\documentclass[lettersize,journal]{IEEEtran}
\usepackage{amsmath,amsfonts}
\usepackage{algorithmic}
\usepackage{algorithm}
\usepackage{array}
\usepackage{textcomp}
\usepackage{stfloats}
\usepackage{url}
\usepackage{verbatim}
\usepackage{graphicx}
\usepackage{cite}
\usepackage{color}
\usepackage{subcaption}
\usepackage{float}
\usepackage{enumitem}
\usepackage{cancel}

\usepackage{xr}
\makeatletter
\newcommand*{\addFileDependency}[1]{
  \typeout{(#1)}
  \@addtofilelist{#1}
  \IfFileExists{#1}{}{\typeout{No file #1.}}
}
\makeatother

\begin{document}

\title{CIRSense: Rethinking WiFi Sensing with Channel Impulse Response
}

\author{Ruiqi~Kong and
        He~Chen 
\thanks{The authors are with the Department of Information Engineering, The Chinese University of Hong Kong, Hong Kong SAR, China. 
(E-mail: \{rqkong, he.chen\}@ie.cuhk.edu.hk).
}
\thanks{Part of the work has been submitted to 2026 IEEE International Conference on Acoustics, Speech, and Signal Processing (ICASSP) \cite{kong2025domino}. 
}
}

\maketitle

\begin{abstract}
WiFi sensing based on channel state information (CSI) collected from commodity WiFi devices has shown great potential across a wide range of applications, including vital sign monitoring and indoor localization. Existing WiFi sensing approaches typically estimate motion information directly from CSI. However, they often overlook the inherent advantages of channel impulse response (CIR), a delay-domain representation that enables more intuitive and principled motion sensing by naturally concentrating motion energy and separating multipath components. 
Motivated by this, we revisit WiFi sensing and introduce CIRSense, a new framework that enhances the performance and interpretability of WiFi sensing with CIR. CIRSense is built upon a new motion model that characterizes fractional delay effects, a fundamental challenge in CIR-based sensing. This theoretical model underpins technical advances for the three challenges in WiFi sensing: hardware distortion compensation, high-resolution distance estimation, and  subcarrier aggregation for extended range sensing. CIRSense, operating with a 160 MHz channel bandwidth, demonstrates versatile sensing capabilities through its dual-mode design, achieving a mean error of approximately 0.25 bpm in respiration monitoring and 0.09 m in distance estimation. Comprehensive evaluations across residential spaces, far-range scenarios, and multi-target settings demonstrate CIRSense's superior performance over state-of-the-art CSI-based baselines.  Notably, at a challenging sensing distance of 20 m, CIRSense achieves at least $3\times$ higher average accuracy with more than $4.5\times$ higher computational efficiency.

\end{abstract}


\section{Introduction}
\IEEEPARstart{W}{ireless} sensing has revolutionized human-computer interaction by enabling non-intrusive monitoring of human activities and vital signs through ubiquitous wireless signals \cite{ma2019wifi}. This technology leverages existing wireless infrastructure to extract subtle environmental changes from radio frequency (RF) signals, offering advantages over traditional sensing approaches that require dedicated sensors or wearable devices \cite{ge2022contactless}. Among various wireless sensing technologies, WiFi sensing stands out for indoor applications due to the ubiquitous presence of WiFi networks, enabling broad coverage without requiring any additional infrastructure \cite{xiao2016survey}. 
Modern WiFi devices provide fine-grained channel state information (CSI) measurements,  which captures the complex frequency response of the channel across OFDM subcarriers. 
The availability of CSI through commercial off-the-shelf (COTS) devices has democratized WiFi sensing research \cite{jiang2021eliminating}. 
Recent advances in CSI-based WiFi sensing have demonstrated promising applications in healthcare monitoring \cite{Zhang2018from,ge2022contactless,liu2021wiphone,fan2024contactless}, human activity recognition \cite{wang2015understanding,chen2018wifi,zou2018deepsense,meneghello2022sharp,yang2023sensefi}, and indoor localization \cite{kotaru2015spotfi,qian2018widar,xie2019mDTrack,zhang2022nlos}, attracting significant attention from both academia and industry.

\begin{figure}
    \centering
    \includegraphics[width=0.9\linewidth]{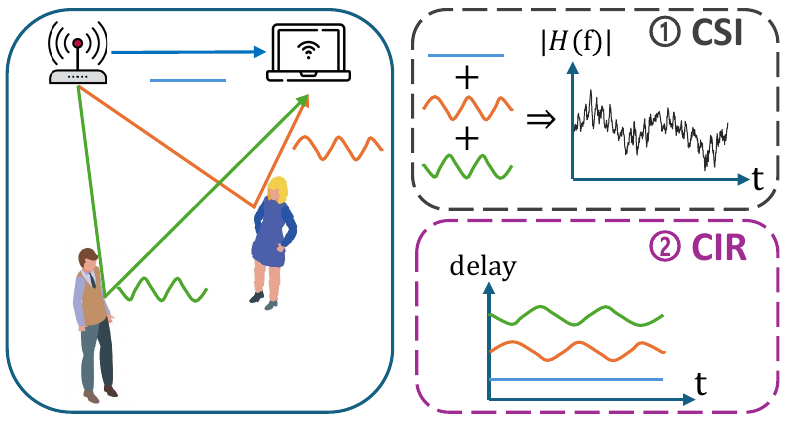}
    \caption{Conceptual comparison of CSI and CIR.}
    \label{fig:motivation}
     \vspace{-0.5cm}
\end{figure}

The fundamental principle underlying WiFi sensing is the detection of temporal variations in wireless propagation paths resulting from interactions with dynamic targets in the environment. Specifically, human activities such as respiration or movement induce subtle changes in the effective path length between the WiFi transmitter and receiver, particularly through reflections involving the human body. These variations manifest as measurable fluctuations in the CSI, thereby enabling the sensing of fine-grained human motion. In addition to CSI’s frequency-domain representation, the wireless channel can also be characterized in the delay domain by the channel impulse response (CIR). Under standard linear time-invariant assumptions, CIR and CSI are Fourier duals, related by the discrete Fourier transform (DFT) and its inverse (IDFT).

The majority of WiFi sensing studies estimate temporal variations in propagation paths directly from the CSI acquired from commodity hardware. \textit{This approach is largely driven by a simplified intuition: given the Fourier duality between CSI and CIR, CSI-based sensing is presumed sufficient, making the conversion to CIR for delay-domain analysis seemingly unnecessary.} In this paper, we challenge this intuition by reexamining WiFi sensing through the lens of CIR. In doing so, we identify that CIR-based sensing offers several advantages over the CSI-based approach prevalent in the literature. First, hardware impairments affect all channel paths in the CIR uniformly, enabling the use of a stable dominant path as a reference to compensate for these impairments. Second, CIR naturally concentrates target motion-induced variations within specific paths rather than dispersing them across all subcarriers in the CSI. This concentration effectively boosts the sensing signal-to-noise ratio (SSNR) for motion extraction, especially for subtle movements like breathing, whose tiny variations are easily buried by noise in CSI. In this regard, CIR-based sensing circumvents the subcarrier aggregation problem inherent in CSI-based methods, where establishing a clear and principled design criterion is often non-trivial. Third, CIR reveals each path’s temporal evolution more distinctly, whereas CSI blends multiple path effects within each subcarrier, as illustrated in \ref{fig:motivation}. This enables selective path processing for different tasks, allowing us to focus processing on the paths most relevant to the sensing objective. Such benefits scale with channel bandwidth (e.g., WiFi 6 and 7 operating at 160 MHz and 320 MHz \cite{cheng2024wifi7}), which enhances path resolution and improves multipath separability. As a result, CIR-based sensing presents a promising direction for overcoming the challenges associated with multi-target WiFi sensing.

Nevertheless, achieving accurate CIR-based sensing is far from straightforward and presents several challenges.
First, the fundamental relationship between variations in CIR taps and the corresponding target motion remains insufficiently characterized in existing research, limiting the interpretability and reliability of CIR-based sensing systems. Second, isolating task-relevant taps requires principled and robust design criteria to distinguish them from numerous multipath components inherent in real-world environments. Third, the limited bandwidth of practical WiFi systems introduces fractional-delay effects, preventing a one-to-one mapping between CIR taps and physical propagation paths. This mismatch complicates the attainment of sub-grid precision, which is essential for high-resolution sensing applications.

To address these challenges, we present CIRSense, the first CIR-based sensing framework to deliver superior performance and computational efficiency with commodity WiFi hardware. CIRSense bridges the gap between the theoretical advantages of CIR-based sensing and practical deployment constraints. Our contributions are threefold:
\begin{itemize}[leftmargin=*]
\item First, we establish a CIR-based sensing model that characterizes the relationship between target motion and delay-domain tap variations, with emphasis on fractional delay effects that are critical for sensing performance. Building on this model, we develop an efficient distortion compensation algorithm (Domino) that exploits the uniform impact of hardware impairments across all CIR paths. By leveraging the inherent stability of the dominant path as a reference, Domino effectively mitigates time-varying hardware distortions while preserving the subtle signal variations induced by target motion. 
\item Second, CIRSense introduces a dynamic path alignment mechanism (Dylign) that harnesses the power concentration property of CIR to address the subcarrier aggregation challenge inherent in CSI-based sensing. Unlike CSI-based methods, which struggle to coherently combine motion signals across widely dispersed subcarriers, Dylign effectively aggregates motion-induced signal energy into a single delay tap, effectively boosting the SSNR. This approach leverages CIR's natural path separation to first identify and focus on motion-relevant taps, avoiding the computational overhead and interference from unrelated paths. Sub-grid precision alignment is then applied to resolve fractional delay effects, delivering both enhanced motion detection for subtle activities like breathing and high-resolution distance estimation of dynamic targets.

\item Third, CIRSense showcases its versatility through a dual-mode operation, supporting both robust respiration monitoring and high-precision target distance estimation. 
Extensive experimental evaluations across diverse environments, including indoor residential spaces, far-range scenarios, and multi-target settings, reveal consistent superiority over state-of-the-art CSI-based methods. 
Furthermore, CIRSense maintains significantly higher computational efficiency, with observed improvements exceeding $4.5\times$, thereby validating the effectiveness of delay-domain signal processing for practical WiFi sensing.
\end{itemize}

We remark that the conference version of this work \cite{kong2025domino} includes only the first contribution listed above, while the second and third contributions are newly introduced in this journal version.

We commit to releasing our code and the real-world dataset to the research community upon acceptance of our work.

\section{Related Work}

\subsection{RF Distortion Compensation}
Existing WiFi sensing schemes experience significant performance degradation or complete failure when deployed on these new-generation cards \cite{enabling24yi}. Specifically, phase-based calibration methods \cite{kotaru2015spotfi, yu2018qgesture, zhu2017phasenoise} address only phase offsets while neglecting magnitude distortions, rendering them incompatible with modern WiFi cards. Reference antenna techniques \cite{zeng2019farsense, indoor17li} cannot address antenna-specific distortions in separate RF chain architectures. Similarly, in the reference subcarrier approach \cite{enabling24yi}, only a subset of subcarrier can be used and it cannot preserve signal integrity for path parameter recovery. Although recent optimization-based methods \cite{meneghello2022sharp} show promise, their substantial computational complexity prevents usage in practical deployments.

\subsection{WiFi-Based Respiration Sensing}
Current WiFi-based respiratory sensing uses CSI magnitude \cite{liu2016Contactless,liu2014wisleep} or phase information \cite{wang2017phasebeat,zhu2017phasenoise,Islam2021BreathTrack} for breathing pattern extraction. However, these methods suffer from ``blind-detection spots" \cite{wang2016matter} where respiratory monitoring fails at certain locations. 
FullBreathe \cite{zeng2018fullbreathe} combines CSI magnitude and phase to reduce blind spots, while learning approaches like ResFi \cite{hu2022resfi} and \cite{fan2024contactless} show promise but require environment-specific training data. 
For extended-range monitoring, FarSense \cite{zeng2019farsense} implements CSI ratio measurements across antennas, EMA \cite{ema} leverages spatial diversity for improved SSNR, and DiverSense \cite{li2022diversense}, the current state-of-the-art, incorporates frequency and time diversity through subcarrier aggregation. However, DiverSense demands high computational resources due to its extensive search spaces for subcarrier processing.

\subsection{WiFi-Based Distance Estimation}

Distance estimation in WiFi systems requires accurate path delay estimation for target-reflected paths. In complex indoor environments, multipath interference obscures target signals, prompting researchers to leverage multi-dimensional information (delay, AoA, Doppler, AoD) for better path separation. SpotFi \cite{kotaru2015spotfi} and WiDeo \cite{joshi2015WiDeo} improve resolution through joint AoA-delay estimation, while mD-Track \cite{xie2019mDTrack} and Widar2.0 \cite{qian2018widar} use all four dimensions for enhanced separation. NLoc \cite{zhang2022nlos} enables NLoS localization but requires hardware capabilities beyond most commercial devices \cite{xu2024pulse,li2024sigcan}. 
SigCan \cite{li2024sigcan} offers an alternative by exploiting multipath signal cancellation in single-target scenarios. However, this approach depends on ideal signal cancellation conditions and linear phase assumption across subcarriers that may not consistently hold.

\subsection{CIR-Based Sensing}
While a handful of prior studies have explored CIR-based sensing, they typically depend on specialized hardware with ultra‑wide bandwidths (e.g., mmWave) or less widely deployed protocols such as IEEE 802.11ay to obtain high‑resolution CIR measurements \cite{multiay23xiong,pegoraro2024jump}. These approaches often simplify by treating CIR tap values as direct proxies for physical path parameters, an assumption that is valid only under ideal conditions with sufficiently high delay resolution. Consequently, their performance hinges on the fine‑grained precision provided by wideband systems (e.g., 1.76 GHz), limiting applicability to widely deployed sub‑6 GHz commodity WiFi devices operating under limited bandwidth constraints. In contrast, CIRSense enables effective CIR‑based sensing on commodity WiFi, delivering sub‑grid precision without specialized hardware or less widely deployed protocols.

\section{Preliminary and Motivations}\label{sec:pre_motivat}
This section first provides an overview of the relationship between CSI, also referred to as channel frequency response (CFR), and CIR. From this foundation, it analyzes the insights from CIR for enhanced sensing. Finally, the remaining challenges associated with CIR-based WiFi sensing are discussed. 

\subsection{Preliminary on CSI and CIR}
\label{sec:pre_cirsense}

\begin{figure*}
  \centering
  \includegraphics[width=0.8\textwidth]{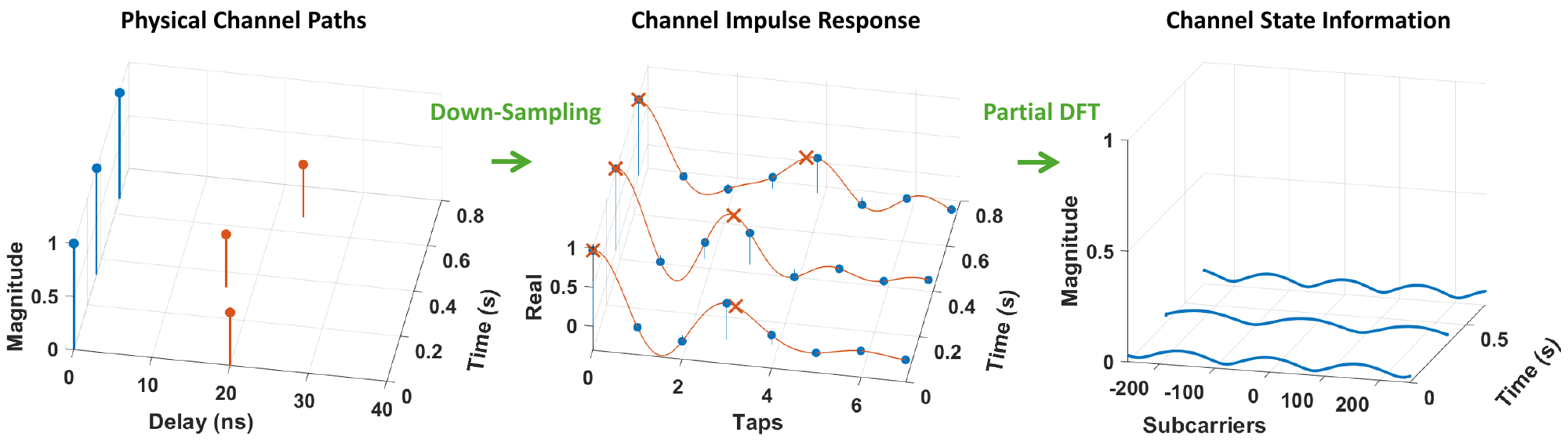}
  \caption{Relationship between physical channel paths, CIR and CSI. }
  \label{fig:pre}
  \vspace{-0.5cm}
\end{figure*}

In wireless communication systems, CSI/CFR and CIR are regarded as two fundamental representations of the wireless channel status. In practical band-limited OFDM systems with bandwidth $B$, the CIR measurement is obtained as a sampled version of the physical channel paths, as shown in Fig.~\ref{fig:pre}. As an example, in the left subfigure of Fig.~\ref{fig:pre}, two physical paths are illustrated: a static path (blue) and a dynamic path (red) over time. 
{The discrete-time CIR measurement for an OFDM system employing an $N$-point DFT can be expressed~as:}
\begin{equation}
\label{tap_model}
   h[n] = \sum_{l=0}^{L-1} \alpha_l e^{-j 2\pi f_c \tau_l} p\left(n T_s - \tau_l \right), 
\end{equation}
where $\alpha_l$ corresponds to the complex channel gain of the $l$-th path, $\tau_l$ represents the time delay of the $l$-th path, $f_c$ denotes the center frequency, $L$ indicates the total number of multipath components, $T_s = \frac{1}{B}$ represents the sampling time interval and $ p(\cdot)$ represents the real-valued pulse shaping function, which is used to shape the transmitted signal in the time domain. 
The discrete-time CIR is illustrated in the middle subfigure of Fig.~\ref{fig:pre}. 
In this subfigure, the red lines correspond to the combined continuous pulse from all propagation paths, where the two red crosses on each line are identified as the ground-truth path delays. 
The blue impulses represent the sampling instants or taps, which are dictated by the system bandwidth.

The sampled CFR/CSI, denoted as $ H[k] $, characterizes the wireless channel's behavior in the frequency domain. Theoretically, CSI and CIR form a unitary DFT pair.
Specifically, the CSI, with a total of $N$ orthogonal subcarriers is expressed as:
\begin{align}
\label{CSI_model_cirsense}
H[k] 
&= \frac{1}{\sqrt{N}} \sum_{l=0}^{L-1}  \alpha_l e^{-j 2\pi (f_c + k \Delta_f)\tau_l} F[k], 
\end{align}
where $\Delta_f$ denotes subcarrier spacing with $\Delta_f = \frac{1}{NT_s}$ and $ F[k] $ represents the frequency response of the pulse shaping function. 
It can be observed that $ H[k] $ provides information about the path gains and delays for all paths at each frequency index $ k $. 
In the middle and right subfigures of Fig.~\ref{fig:pre}, the corresponding sampled CIR and CSI are illustrated for the channel path situations shown in the left subfigure. 
The bandwidth is set to 160 MHz, and the parameter 
$N=512$ are adopted in accordance with the IEEE 802.11ac standard \cite{80211ac}. In practice, WiFi standards use only a subset of subcarriers for data transmission. This partial usage creates trouble when converting CSI to CIR because the standard IDFT cannot be directly applied to recover accurate CIR from incomplete CSI measurements. We address this issue in Section~\ref{cir_measure}. 

We remark that the limited system bandwidth results in a discrepancy where CIR taps do not perfectly align with actual physical propagation paths. For clarity, we adopt the following terminologies hereafter: \textit{path parameters/coefficients} refer to the sub-grid estimates obtained through CIRSense as detailed in Section~\ref{method_cirsense}, while \textit{tap values} denote the grid-level values in the CIR directly recovered from CSI. 

\subsection{Motivations of CIRSense}
\label{motivation_cirsense}

The CIR does not inherently provide additional information compared to CSI, as they are merely different domain representations of the same channel characteristics. However, delay-domain analysis offers more direct and physically intuitive insights for sensing algorithm design.

\textbf{First, the delay domain uniquely enables selective path information recovery without costly full channel decomposition.} As illustrated in Fig.~\ref{fig:pre}, CIR provides an intuitive approximation to the physical channel paths, revealing a critical insight: we can directly identify and process only the paths relevant to our sensing tasks. The CIR representation naturally separates multipath components along the delay axis, allowing us to selectively focus computational resources on specific taps of interest, the dominant static path closely aligns with the tap exhibiting the strongest magnitude, while the motion-related dynamic path corresponds to the tap with maximal temporal variance, as experimentally demonstrated in Appendix~\ref{pre_exp}. This selective processing capability reduces computational complexity and accumulated estimation errors inherent in traditional joint parameter estimation problems, particularly in multipath-rich environments. 
This advantage becomes particularly valuable when addressing two fundamental challenges in WiFi sensing: hardware-induced distortions and precise target distance estimation, as we demonstrate in detail in Section~\ref{method_cirsense}.

\textbf{Second, when sensing signal-to-noise ratio (SSNR) is concerned, it is observed that in CIR, the majority of the power of the dynamic path is more concentrated than in CSI.} Here, SSNR, as defined in~\cite{ema}, is used to quantify sensing signal quality. It is defined as the ratio of the target-reflected signal power (i.e., the power of the dynamic path) to the noise power. This metric reveals a fundamental advantage of the delay domain representation: whereas in CSI, signal power becomes dispersed across all subcarriers due to the properties of the unitary DFT, the CIR representation naturally concentrates the target-induced signal power within a limited number of delay taps. As mathematically demonstrated by the $\frac{1}{\sqrt{N}}$ factor in Equation~\ref{CSI_model_cirsense} and visually confirmed in Fig.~\ref{fig:pre}, the signal power corresponding to a single propagation path undergoes significant dispersion across all subcarriers in the frequency domain, while remaining localized in the delay domain. This inherent power concentration in CIR potentially enhances both the accuracy and robustness of underlying sensing tasks by improving the effective SSNR, as experimentally demonstrated in Appendix~\ref{pre_exp}. 
Moreover, this perspective sheds light on why subcarrier aggregation techniques for SSNR maximization in CSI-based approaches \cite{zeng2019farsense, li2022diversense} often lack a clear, principled design criterion. Their core aim can be reframed as concentrating dispersed signal power into a coherent representation, conceptually equivalent to concentrating power into a single tap, as derived in Section~\ref{impact_fd}. Viewed this way, CIR offers a principled foundation for algorithm design, enabling reduced algorithmic complexity alongside improved sensing performance. Our experimental results in Section~\ref{evaluation} empirically validate the superior performance of CIRSense compared to prior CSI-based baselines \cite{zeng2019farsense, li2022diversense}.

\subsection{Challenges of CIRSense}

Despite its advantages, CIR-based WiFi sensing faces several technical challenges that must be resolved. 
First, the discrete nature of CIR taps introduces fundamental limitations in representing physical paths. Fractional delays, the misalignment between actual propagation delays and sampling instants, largely impact sensing performance. Existing CSI-based motion models fail to account for these critical delay-domain phenomena, limiting their ability to reliably interpret motions from CIR variations. Second, as established in Sections~\ref{motivation_cirsense}, the three major challenges (i.e., distortion compensation, distance estimation, SSNR maximization) in existing WiFi sensing require effective solutions to accurately estimate parameters of task-related paths. While CIR's energy concentration property offers advantages for selective tap processing, challenges remain in realizing its full potential for WiFi sensing, particularly in the transition from raw tap values to precise path parameters. Most notably, the relationship between tap values and actual path parameters becomes nonlinear due to fractional delay effects. This challenge necessitates the development of new algorithms to accurately recover path parameters from task-relevant taps.

In the subsequent sections, the challenges associated with CIR-based sensing are systematically addressed. First, a CIR-based sensing model incorporating fractional delay analysis is introduced in Section~\ref{cir_model}. 
Second, Section~\ref{method_cirsense} introduces a new parameter estimation framework that simultaneously addresses requirements for RF distortion compensation, fine-grained distance estimation, and SSNR maximization.

\section{CIR-based Sensing Model}
\label{cir_model}

In this section, the motion model for CIR-based sensing is first introduced to establish the relationship between variations in CIR taps and target motion. 
Following this, we discuss the major factor that fundamentally limits sensing performance.

\subsection{ Motion Model }

Propagation paths in a wireless environment can be categorized into static paths and dynamic paths \cite{wang2015understanding}. Static paths remain constant over time $ t $, while dynamic paths vary with target movements. We begin with a typical single-target WiFi sensing scenario, in which a pair of WiFi transceivers is positioned at fixed locations. The dynamic path corresponds to the signal reflected by the moving target. By aggregating the contributions from static paths into a constant complex value for each tap, the CIR can be expressed as:
\begin{equation}
\label{motion_model}
    h[n,t] = h_s[n] + \alpha[t] e^{-j 2\pi f_c \tau[t]} p[n,\tau[t]] + z[n,t],
\end{equation}
where $ h_s[n] $ represents the aggregated component of all static paths, $ \alpha[t] $ and $ \tau[t] $ denote the time-varying complex path gain and delay of the dynamic path, respectively, $p[n,\tau[t]]$ is defined as $p\left(n T_s - \tau_l\right)$, representing the sampled pulse shaping function, and $z[n,t] $ is additive white Gaussian noise. For multi-target scenarios, this model can be readily extended by incorporating additional dynamic components.

\begin{figure}
  \centering
  \subfloat[]{\includegraphics[width=0.9\linewidth]{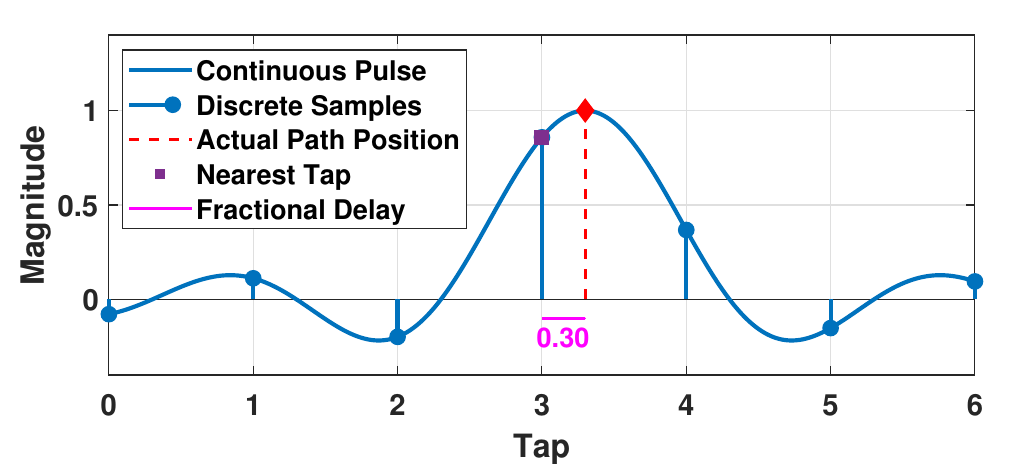}\label{pulse}}
  
  \subfloat[]{\includegraphics[width=0.52\linewidth]{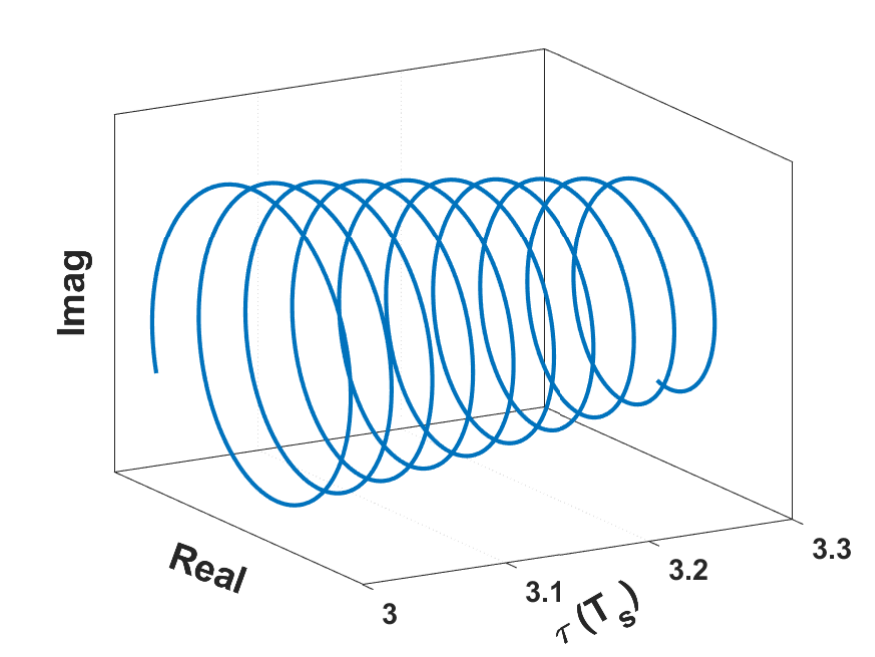}\label{iqplane_3d}}
  \subfloat[]{\includegraphics[width=0.43\linewidth]{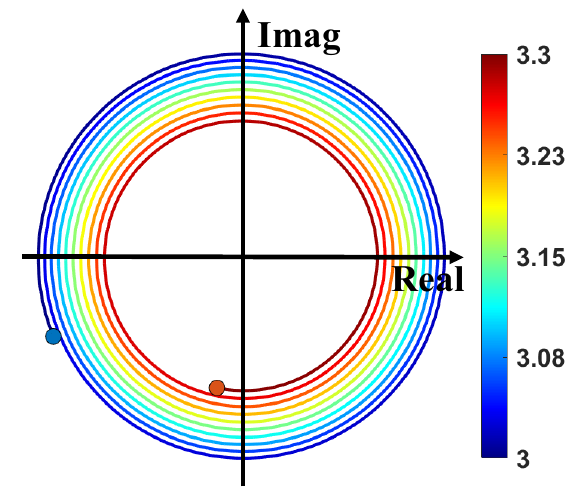}\label{iqplane}}
\caption{CIR-based motion model. (a) Sampling of the pulse shaping function with fractional delay. (b) Target tap value varies with delay changes. (c) Complex plane representation.}
\label{fig:motion_model}
 \vspace{-0.5cm}
\end{figure}
   
Fig.~\ref{fig:motion_model} provides an intuitive visualization of how the CIR varies with target locations and movements, excluding the effect of noise. The biggest difference between the CIR-based and the CSI-based sensing model is the pulse-shaping effect, as discussed in the following. 

First, the magnitude of the taps in the CIR is governed by two principal factors: path attenuation $ \alpha[t] $ and pulse-shaping value $ p[n, \tau[t]] $. 
While attenuation, which follows an inverse relationship with distance, changes slowly with target displacements, the pulse-shaping function exerts a considerably more pronounced influence on magnitude due to fractional delay. The fractional delay $ \Delta(\tau) $ arises due to the misalignment between the physical delay $ \tau $ and the discrete sampling grid $ nT_s $. The fractional delay $\Delta(\tau)$, defined as the residual between the physical propagation delay $\tau$ and its nearest integer multiple of the sampling interval $T_s$, can be expressed as:
$$\Delta(\tau) = \tau - {\rm{round}}(\frac{\tau}{T_s})\cdot T_s,$$ 
where ${\rm{round}}(\cdot) \mathop: \mathbb{R} \to \mathbb{Z}$ represents the function that rounds each real number to the nearest integer. 
As shown in Fig.~\ref{pulse}, one can easily observe that the magnitude of the nearest tap decreases with increasing fractional delay.

Second, phase variations are solely caused by the term $e^{-j 2\pi f_c \tau[t]}$, as the pulse shaping function is a real-valued function that only influences magnitude. Consequently, the phase of the dynamic component rotates clockwise in the complex plane as $\tau[t]$ increases. Furthermore, due to the large value of the carrier frequency $f_c$, even small changes in $\tau$ can result in significant phase variations. For example, when $f_c = 5.25$ GHz, a displacement of 5.7 cm results in a $2\pi$ phase shift. Fig.~\ref{iqplane_3d} and ~\ref{iqplane} demonstrates the combined effect of distance displacement on the complex plane of tap $h[3]$, where $f_c = 5.25$ GHz, $B = 160$ MHz, and $\tau[t]$ changes from $3T_s$ to $3.3T_s$. Specifically, the center of the circular trajectory corresponding to the aggregated static component $h_s[n]$. 

It is important to note that the above observations are derived without making any assumptions about the specific pulse shaping filter used, indicating their broad applicability across various filter types. We validated the proposed motion model through controlled experiments in Appendix~\ref{modelvalidation}. 

\subsection{Impact of Fractional Delay}
\label{impact_fd}
In this section, we discuss the impact of fractional delay on path parameter recovery and respiration pattern extraction. 
First, fractional delay presents a major challenge in path parameter recovery, as simple tap approximation methods become inadequate. We note that the potential path delay estimation error introduced by fractional delay can reach up to $0.5T_s\times c = 0.9375$ meters in 160MHz systems, where $c$ represents the speed of light. This magnitude of error is unacceptable for high-precision sensing applications, necessitating advanced estimation techniques.

For respiration pattern extraction, based on the analysis in \cite{ema}, SSNR constitutes a critical parameter that directly influences both sensing accuracy and effective range. Following that, we now examine how fractional delay impacts SSNR. During normal breathing, chest displacement typically ranges from 5 mm to 12 mm \cite{lowanichkiattikul2016impact}. Within this limited displacement range, $ p[n, \tau[t]] $ exhibits negligible variation. These minimal fluctuations validate the approximation of $p[n, \tau[t]]$ as a constant value $\bar{p}[n, \tau]$ for respiratory monitoring applications. 
Similarly, this small displacement has negligible impact on the dynamic path gain $ \alpha[t] $, allowing it to be treated as a constant value \cite{faria2005modeling}. Given these considerations, the SSNR for CIR signal can be expressed as:
\begin{align}
\label{eq:ssnr}
    \gamma &= \frac{P_{target}}{P_{noise}} = \frac{\mathbb{E}(|h_d[n,t]|^2)}{\mathbb{E}(|z[n,t]|^2)} 
    = \frac{|\alpha \bar{p}[n, \tau]|^2}{\sigma^2[n]},
\end{align}
where $ \mathbb{E} $ denotes the expectation operation, $ h_d[n,t] $ represents the dynamic component corresponding to the target-reflected signal, and $ \bar{p}[n, \tau]$ represents the mean value of $p[n, \tau[t]]$. 
It is observed that under perfect sampling conditions without fractional delay, where $ \bar{p}[n, \tau] \approx 1$, the dynamic path power is entirely concentrated in a single tap, resulting in the maximum SSNR $ \gamma = \frac{|\alpha|^2}{\sigma^2} $. However, as the fractional delay $ \Delta(\tau) $ increases, the SSNR decreases to $ \gamma = \frac{|\alpha p[\Delta(\tau)]|^2}{\sigma^2} $. Note that even in the worst case, the SSNR in CIR is much larger than that of any subcarrier in the CSI, given by $ \gamma_{\rm{sc}} = \frac{|\alpha|^2}{N\sigma^2} $, where the least number of subcarriers $N$ used in practical WiFi systems is 64 in 20MHz systems.

In the following section, we propose approaches to mitigate the impact of fractional delay on these sensing tasks, thereby enhancing overall sensing performance and reliability.

\section{CIR-based Sensing Mechanisms}
\label{method_cirsense}

This section presents mechanisms designed to accurately obtain the CIR from CSI measurements collected using commodity WiFi systems. Building on this, we develop a compensation method to address time-varying RF distortions commonly observed in practical CSI measurements. This section further presents methods designed to minimize distance estimation errors and maximize the SSNR. 

\subsection{CIR from Commodity WiFi}
\label{cir_measure}

The direct application of CIR-based sensing methods to commodity WiFi systems faces a fundamental implementation issue: obtaining accurate CIR estimations. In practical WiFi systems, CIR is not directly available as an intermediate output and must be derived from CSI measurements, which only contain values from a subset of subcarriers. A direct IDFT approach to obtain CIR from these incomplete CSI measurements leads to inaccurate results. 
Due to the inherent sparsity of wireless channels, the number of taps in the delay domain is often significantly smaller than the DFT length $ N $. Consequently, the least-squares (LS) method can be employed to estimate the CIR from CSI measurements with partial subcarrier usage \cite{ls,kong2024csirff}. 
The LS method estimates the CIR $\mathbf{h}$ as follows:
\begin{align}
\label{ls_cirsense}
    \hat{\mathbf{h}} &= (\mathbf{F}_{\mathcal{K},\mathcal{L}}^H \mathbf{F}_{\mathcal{K},\mathcal{L}})^{-1} \mathbf{F}_{\mathcal{K},\mathcal{L}}^H \mathbf{H} , 
\end{align}
where $\mathbf{F}$ represents the full unitary DFT matrix, and $\mathbf{F}_{\mathcal{K},\mathcal{L}}$ is the sub-matrix of $\mathbf{F}$, comprising all rows corresponding to the active subcarrier set $\mathcal{K}$ and all columns corresponding to the potential tap set $\mathcal{L}$. The operators $(\cdot)^H$ and $(\cdot)^{-1}$ denote the Hermitian transpose and matrix inverse, respectively. A simulation-based validation is provided in Appendix~\ref{cir_acq_val}.

\subsection{Principle of Path Alignment}
\label{path_align}

To address the negative impact of fractional delay, a delay-shifting operation can be applied to the CIR to precisely align the target path with a discrete CIR tap. The delay-shifted CIR, denoted as $ h[n + \Delta'] $, is expressed as:
\begin{align}
    h[n + \Delta'] &= \sum_{l=0}^{L-1} \alpha_l e^{-j 2\pi f_c \tau_l} p\left( (n + \Delta')T_s - \tau_l \right) \nonumber
\end{align}
where $ \Delta' $ represents the delay shift applied to all taps. When $ \Delta' =\frac{\Delta(\tau_l)}{Ts} $, this operation aligns the physical delay $\tau_l$ to its closest CIR tap, effectively compensating for the fractional delay of the desired path. {Note that $ \Delta' $ may take fractional values and the expression $h[n + \Delta']$ is misused for notation simplicity.} The delay-shifting operation can be implemented efficiently in the frequency domain by applying phase shifts to the CSI measurements, see more implementation detail in Appendix~\ref{principle_fractionaldelaycom}. 
The remaining challenges involve identifying relevant paths for different tasks and determining the fractional delay for compensation.

\subsection{Dominant Path Alignment for RF Distortion Compensation}
\label{rfdistortion_cirsense}

Reliable WiFi sensing critically depends on high-fidelity sensing signals, which are largely affected by time-varying RF distortions in commercial devices. These hardware-induced distortions obscure the subtle CSI variations caused by target motion \cite{terry2002ofdm}. Existing compensation approaches often rely on strong hardware assumptions or sacrifice signal integrity \cite{enabling24yi}. 
Our proposed solution to this issue is based on the insight that the magnitude and phase variations caused by hardware imperfections are consistent across all paths within each CSI/CIR. Therefore, the dominant static path can be selected as a reference to align other paths for distortion compensation. In practical deployments, static dominant path configurations can be readily established, for example, through monostatic setups where a single router performs both communication and sensing \cite{huaweicsisensing}, or by strategically positioning the transmitter and receiver to preserve a stable dominant path, such as mounting them near the ceiling to avoid potential dynamic blockage. 

The undesired phase offset contains multiple contributions \cite{zhu2018pi}. Some of them, including the channel frequency offset (CFO), the phase-locked loop (PPO) and the phase ambiguity (PA), although changing in time, have the same value across the paths associated with each receiving antenna. The sampling frequency offset (SFO) and packet detection delay (PDD) are instead a common delay shift for each path. 
In addition, AGC induces a random gain in each measurement. 
The noise-free CIR of the $n$-th tap at a receiving antenna, accounting for hardware distortions, can be expressed as
$$
h[n] = \beta e^{-j\theta}\sum_{l=0}^{L-1} \alpha_l e^{-j 2\pi f_c \tau_l} p[n, \tau_l + \epsilon],
$$
where $\beta$ represents the magnitude distortion, $\theta$ represents the phase offset common for each path, and $\epsilon$ denotes the common delay shift.

Based on these insights, we propose a two-step solution to eliminate hardware-induced distortions for recovering clean motion signals. 
The first step addresses the random delay $\epsilon$ by aligning the first tap $h[0]$ with the strongest propagation path (dominant path). 
In typical indoor environments, the path exhibiting maximum gain $|\alpha|$ generally corresponds to a static path, characterized by delay $\tau_0$ \cite{meneghello2022sharp}. It typically exhibits higher power compared to other paths. This property, combined with the rapid decay of the pulse shape function $p[n, \tau_l + \epsilon]$ for non-zero differences between $nT_s$ and $\tau_l + \epsilon$, permits the derivation of the following approximation:
\begin{align}
    h[n_0]  &\approx \beta e^{-j\theta}\alpha_0 e^{-j 2\pi f_c \tau_0} p[n_0, \tau_0 + \epsilon] ,
\end{align}
where $n_0$ is the closest tap to $\tau_0+\epsilon$ and $h[n_0]$ is value of the strongest tap. 
The optimal delay shift parameter $\epsilon'_{\text{est}}$ can be determined through the following maximization:
\begin{align}
\label{losalign}
\epsilon'_{\text{est}} &= \arg\max_{\epsilon'} |h[0 + \epsilon']| \nonumber\\
&\approx \arg\max_{\epsilon'} |\beta e^{-j\theta}\alpha_0 e^{-j 2\pi f_c \tau_0} p[0+\epsilon', \tau_0 + \epsilon]| ,
\end{align}
where the objective function reaches its optimum at $\epsilon' = -\frac{\tau_0+\epsilon}{T_s}$. 
This optimization effectively shifts the CIR to maximize the power concentration in the first tap. After applying this delay shift, the aligned CIR can be expressed~as: 
$$h[n+\epsilon'_{est}] = \beta e^{-j\theta}\sum_{l=0}^{L-1} \alpha_l e^{-j 2\pi f_c \tau_l} p[n, \tau_l - \tau_0].$$
As the second step, the remaining hardware-induced distortions can be eliminated by:
\begin{align}
\label{rfcom}
h'[n] &= \frac{h[n+\epsilon'_{est}]}{h[0+\epsilon'_{est}]} \approx \frac{ \cancel{\beta e^{-j\theta}} \sum_{l=0}^{L-1} \alpha_l e^{-j 2\pi f_c \tau_l} p[n, \tau_l - \tau_0]}{ \cancel{\beta e^{-j\theta}} \alpha_0 e^{-j 2\pi f_c \tau_0}} \nonumber\\
&= \sum_{l=0}^{L-1} \alpha'_l e^{-j 2\pi f_c \tau'_l} p[n, \tau'_l] ,
\end{align}
where $h'[n]$ represent the clean version of CIR after distortion mitigation, $\alpha'_l = \frac{\alpha_l}{\alpha_0}$, $\tau'_l = \tau_l-\tau_0$. 
This ratio effectively cancels out the time-varying hardware distortion terms $\beta$ and $\theta$. Since the channel parameters of the strongest static path remain constant over time, any variations in the resulting normalized CIR can be attributed purely to target motion, free from distortions. We refer to our proposed compensation method as Domino throughout subsequent sections.

\subsection{Dynamic Path Alignment for Enhanced Sensing}
\label{targetpa}

To achieve a fine-grained distance estimation and maximize SSNR, the goal is to estimate the factional delay $\Delta(\tau)$ for the target-reflected path and align this dynamic path with a sampling tap. This alignment methodology builds on the key observation that among all CIR taps, the one exhibiting maximal temporal variance corresponds most closely to the actual physical path delay of the target path. 

To verify this principle, we analytically derive the variance characteristics of cleaned CIR $ h'[n,t] $ under single-target respiratory motion conditions. First, it is noted that $ h'_s[n] $, $ \alpha' $, and $ \bar{p}[n, \tau'] $ are constant over time as established in Section~\ref{cir_model}. The mean of $ h'[n,t] $ is given by 
$$ \mathbb{E}[h'[n,t]] = h'_s[n] + \alpha' \bar{p}[n, \tau'] \mu_{\tau'}, $$ 
where $ \mu_{\tau'} = \mathbb{E}[e^{-j 2\pi f_c \tau'[t]}] $ represents the mean of the complex exponential term. The variance can be expressed as 
\begin{equation}
\label{var_cir}
   \text{Var}(h'[n,t]) = |\alpha' \bar{p}[n, \tau']|^2 (1 - |\mu_{\tau'}|^2).
\end{equation}
Here, $ |\mu_{\tau'}|^2 $ and $ |\alpha'|^2 $ are identical for all taps $ n $. According to the properties of the pulse shaping function, $\bar{p}[n, \tau'] $ is maximized when $ |\tau' - nT_s| $ is minimized. Consequently, the tap with the largest variance corresponds to the tap closest to the physical path delay. Appendix~\ref{pre_exp} further validates this insight experimentally.

By iteratively searching delay shift $ \Delta' $, the fractional delay can be minimized, leading to improved alignment of the CIR with the dynamic path. This method not only enhances the accuracy of delay estimation but also maximizes the SSNR by concentrating the signal power in a single tap. To find the optimal delay shift $ \Delta' $, the previous insight is leveraged to formulate an optimization problem. Specifically, the tap with the largest variance is identified as the closest tap to the physical path, denoted as $n^*$, and the fractional delay is estimated by minimizing the misalignment between the tap $n^*$ and the physical path delay. The optimization problem can be mathematically formulated~as
\begin{equation}
\label{rd_est}
    \Delta'_{\text{opt}} = \arg\max_{\Delta'} \text{Var}\left(h'[n^* + \Delta', t]\right),
\end{equation}
where $ \text{Var}\left(h'[n^* + \Delta', t]\right) $ represents the variance of the delay-shifted CIR tap $n^*$. It is noted that, in Equation~\ref{var_cir}, the terms $ |\mu_{\tau'}|^2 $ and $ |\alpha'|^2 $ are independent of the channel tap index $ n $. This implies that the delay-shifting operation does not affect their values. Consequently, the variance of the time-shifted CIR, $ \text{Var}(h'[n + \Delta', t]) $, is linearly proportional to $ |\bar{p}[n^* + \Delta', \tau']|^2 $. As $ \Delta' $ is adjusted, the variance reaches its maximum when $ \Delta' = \frac{\Delta(\tau')}{Ts} $. At this point, the delay-shifted CIR is optimally aligned with the delay of the dynamic path, and the variance is maximized. 

This delay-shifting operation is evaluated over a range of potential fractional delays $ \Delta' $, and the value that maximizes the variance is selected as the optimal delay shift. The search space for $ \Delta' $ is typically constrained to $ |\Delta'| \leq \frac{1}{2} $. To further improve computational efficiency, a coarse-to-fine search strategy can be employed. Initially, a coarse search is performed over a wide range of $ \Delta' $ to identify a rough estimate of the optimal delay shift. This is followed by a fine search in the vicinity of the coarse estimate to refine the result. This two-step approach reduces the computational complexity while maintaining high accuracy in fractional delay estimation.

{CIRSense then uses the temporal characteristics of the aligned dynamic path for respiratory monitoring.} Specifically, the shifted tap exhibiting maximum variance $h'[n^* + \Delta'_{\text{opt}}, t]$ serves as the primary signal for respiration rate estimation. 
For target distance estimation tasks, additional processing is required to convert path delay measurements into absolute distance values. Following the static path alignment procedure, all subsequent delay measurements are referenced relative to the LoS path delay, expressed as $\tau'_l = \tau_l-\tau_0$ in Equation~\ref{rfcom}. After identifying the optimal time shift $\Delta'_{\text{opt}}$ in Equation~\ref{rd_est} during dynamic path alignment, the relative delay of the target can be computed as $\tau_{\text{est}}=(n^*+\Delta'_{\text{opt}})\times T_s$. This relative delay is then converted to the absolute path length through $d_{\text{target}} = c \cdot \tau_{\text{est}} + d_0$,
where $d_0$ represents the known transceiver separation distance corresponding to the strongest static path. 
The transceiver setup is a common known prior in existing distance estimation algorithms, as absolute path delay is hard to obtain due to RF distortions.  
It's important to note that while distance estimation requires this information, respiration sensing based on temporal pattern can operate without such knowledge. We refer to our proposed enhanced sensing method as Dylign throughout subsequent sections.

\begin{figure}
    \centering    \includegraphics[width=\linewidth]{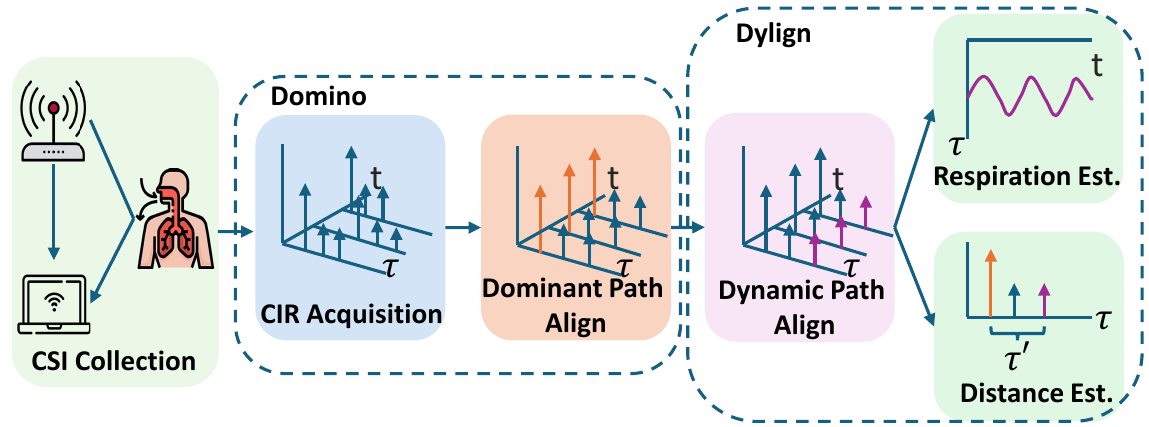}
    \caption{The Signal Processing Workflow in CIRSense.}
    \label{fig:framework}
     \vspace{-0.5cm}
\end{figure}

\subsection{CIRSense Workflow and Validation}

To summarize, Fig.~\ref{fig:framework} depicts the complete signal processing workflow in CIRSense, which consists of three key processing stages and two estimation outputs. The system begins with CSI collection, where WiFi signals are transmitted between a pair of transceivers and interact with the human subject. 
The collected CSI measurements then undergo processing described in Section~\ref{cir_measure} to obtain CIR estimations. This is followed by a two-stage alignment process. 
{First, individual CSI measurements undergo dominant path alignment to compensate for RF distortions, following the methodology described in Section~\ref{rfdistortion_cirsense}. Subsequently, dynamic path alignment is applied to one set of cleaned CIR estimates to improve distance estimation accuracy while enhancing SSNR, as detailed in Section~\ref{targetpa}. At this stage, CIRSense can output the estimated distance and respiration pattern of the target.}

\begin{figure}
  \centering
  \subfloat[RF distortions elimination.]{\includegraphics[width=0.5\linewidth]{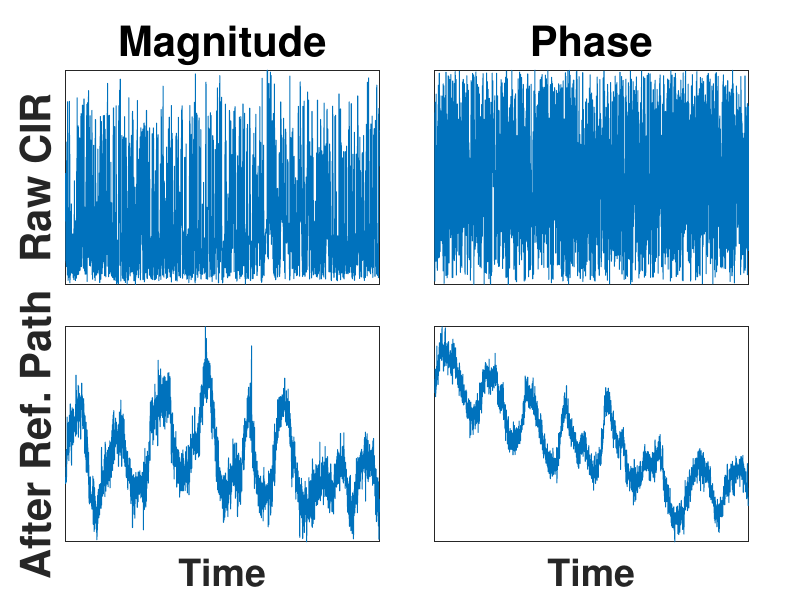}\label{com_dis}}
  \subfloat[SSNR improvement.]{\includegraphics[width=0.5\linewidth]{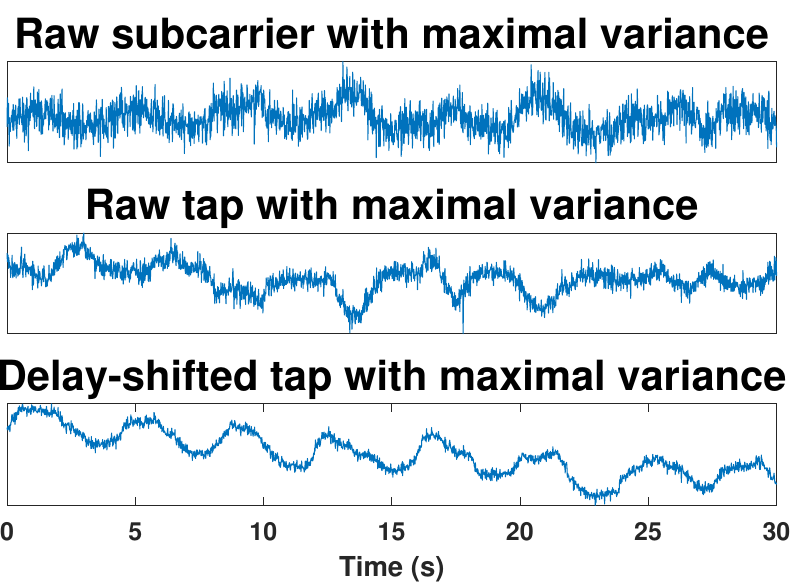}\label{fig:ssnr_val}}
  
  \subfloat[Fine-grained path delay estimation.]{\includegraphics[width=0.95\linewidth]{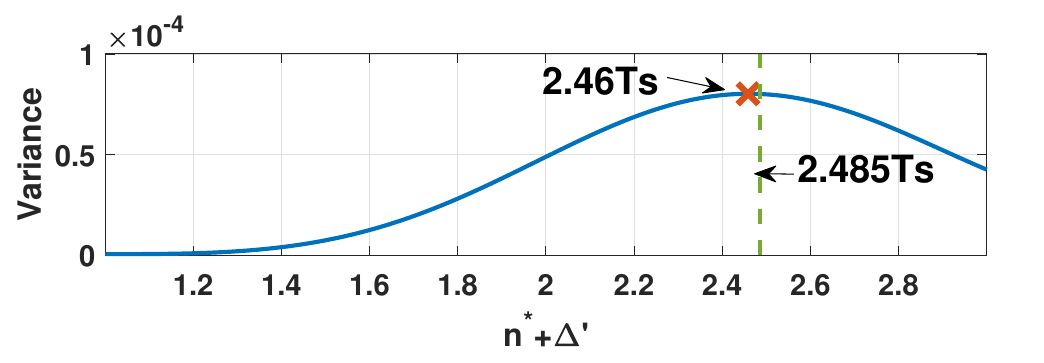}\label{fig:tof_est_val}}
\caption{Experimental validation of CIRSense. } 
 \vspace{-0.5cm}
   \end{figure}
   
The effectiveness of CIRSense is now experimentally validated. In this experiment, the WiFi transceiver is configured to operate at 5.25GHz with a bandwidth of 160MHz. The transmitter and receiver are positioned with a 60 cm separation, while a test person is located 2.63 m away to the transceiver. This configuration is designed to produce maximal fractional delay conditions, representing the worst scenario for distance estimation accuracy. Respiratory motion is captured for about 30 seconds.
   
\textbf{Distortion Compensation:} 
As we can see from the first row of Fig.~\ref{com_dis}, the tap's magnitude and phase suffer from significant time-varying distortions. We then suppress these distortions by our static path alignment method. As evidenced in second row of Fig.~\ref{com_dis}, our CIR-based method enables effective RF distortions compensation while preserving periodic breathing pattern.

\textbf{SSNR Maximization:} 
A direct comparison of SSNR is presented in Fig.~\ref{fig:ssnr_val}, contrasting the highest-variance CSI subcarrier with the highest-variance CIR tap. 
The results clearly show that periodic respiratory patterns are more distinctly observable in CIR, as CSI inherently distributes signal power across subcarriers. Further, after alignment with the dynamic path, the SSNR is further increased with more clear patterns. This improvement is attributed to the concentration of the signal power in a single tap. 

\textbf{Distance Estimation:}  
In this experimental setup, the ground-truth relative delay of $2.485T_s$ is indicated by the green dashed line in Fig.~\ref{fig:tof_est_val}. The results demonstrate the efficacy of the dynamic path alignment operation, with the variance reaching its peak at 
$ \tau_{\text{est}} = (n^*+\Delta'_{\text{opt}})\times T_s=2.46T_s $, resulting in an estimation error of $ (2.485 - 2.46)T_s = 0.025T_s $. For a 160MHz system, this error translates to a path length discrepancy of 0.02 m, indicating the high precision of the proposed distance estimation method.

\section{Systematic Evaluation}
\label{evaluation}
The distance estimation and respiration rate estimation performance of CIRSense is now systematically evaluated and compared with state-of-the-art methods.  

\subsection{Experimental Settings}
\label{sec:exp_setting}

To comprehensively evaluate the sensing capability of CIRSense, we conducted a series of experiments across diverse environmental conditions, including line-of-sight (LoS) scenarios at varying sensing ranges, non-line-of-sight (NLoS) scenarios, and multi-target settings. These experiments were designed to systematically assess robustness and accuracy of CIRSense under real-world constraints.

\textbf{Baselines:} A comparative analysis is conducted among CIRSense, and two state-of-the-art CSI-based respiration monitoring techniques, FarSense \cite{zeng2019farsense} and DiverSense \cite{li2022diversense}. 
These two baselines aggregate sensing signals from multiple subcarriers to improve sensing performance. 
For distance estimation, SigCan \cite{li2024sigcan} serves as the baseline, representing the current state-of-the-art method for CSI-based moving target distance estimation without heavily relying on hardware synchronization of multiple receiving antennas, an assumption proven impractical in new generation WiFi cards \cite{enabling24yi}. {It is worth noting  that none of these baselines can achieve dual-mode sensing capabilities of CIRSense, as they were designed and optimized for a single sensing task, either respiration monitoring or distance estimation.}

After the CSI time-series data is acquired, the method described in Section~\ref{cir_measure} and \ref{rfdistortion_cirsense} is applied to compensate for time-varying RF distortions and obtain cleaned CIR. 
It should be noted that the three baseline methods \cite{zeng2019farsense,li2022diversense,li2024sigcan} originally use the CSI ratio from the Intel 5300 NIC as algorithm input. 
However, prior research \cite{meneghello2022sharp,enabling24yi} and our experimental results consistently demonstrate that the CSI ratio lacks robustness when receiver chains exhibit different distortions. 
To ensure a fair comparison across all methods, the same distortion compensation method proposed in this work is applied to all three baselines. 
The performance degradation associated with the use of the CSI ratio is further analyzed in Section~\ref{impact:dis_com}. 
Following RF distortion compensation, a moving average filter is employed to smooth the signals in the time domain, mitigating noise and enhancing underlying patterns. 
Subsequently, different techniques are used for respiration signal extraction or distance estimation.

\textbf{Data Collection:} The data collection process is repeated multiple times on different days. All CSI measurements are collected using the Picoscenes platform \cite{jiang2021eliminating} on a mini-PC equipped with an Intel AX200 network interface card (NIC) with a single receiving antenna. The sampling rate is set to 500 Hz for distance estimation, aligning with the baseline method to ensure fair comparisons. For respiration rate estimation, a sampling rate of 200 Hz is used, consistent with the baseline methods. The transmitter is a modified ASUS TUF Gaming AX3000 router that transmit using a single antenna. The network is configured to operate at 5.25 GHz with a bandwidth of 160 MHz, using the 802.11ax protocol. The ground truth of the respiration rate is collected using a commercial Neulog respiration belt \cite{Belt}. Note that our experiments employ distinct human postures for different measurement objectives. For distance estimation, the target maintains a standing position while breathing normally, ensuring consistent body alignment along the measurement axis. Conversely, for respiration rate estimation, the target adopts a seated position to minimize whole-body movements that could interfere with the detection of respiratory motion patterns.

\textbf{Parameter configurations:} The potential tap set $\mathcal{L}$ in Equation~\ref{ls_cirsense} is determined by the expected delay range and pulse shaping leakage \cite{bala2013shaping,xu2024pulse}. Given the rapid decay characteristic of the impulse response of the pulse shaping filter, it is safe to assume that the pulse shape will decay to zero after 20 sampling time intervals. Further, considering the maximal path length with noticeable path gain is less than 60 m \cite{delay2022}, we establish an upper bound of 30 taps, corresponding to a maximum distance of 56.25 m ($30\times T_s \times c$) in 160 MHz systems. To accommodate pulse shaping leakage while ensuring comprehensive coverage, we define the potential tap set as 
$\mathcal{L} =\{-20,-19,...,50\}$ in our experiments. 
For path alignment method, an iterative search procedure is employed to identify the optimal fractional delay estimate within the specified tap range. The search process consists of two stages: an initial coarse-grid search with a precision of $T_s/20$, followed by a refined search with a higher precision of $T_s/200$. This two-stage approach is consistently applied throughout all subsequent performance evaluations.

\textbf{Metrics:} 
The performance of the proposed method is evaluated using the following metrics: (a) distance estimation error: the absolute differences between the estimated distance from the target to the transceiver and the ground truth distances; (b) respiration rate error: the absolute differences between the estimated respiration rate and the ground truth respiration rates, measured in breaths per minute (bpm).

\subsection{Sensing Range Evaluation in LoS Settings}
\label{sslos}

\begin{figure*}
  \centering
  \subfloat[Indoor living room.]{\includegraphics[width=0.4\linewidth]{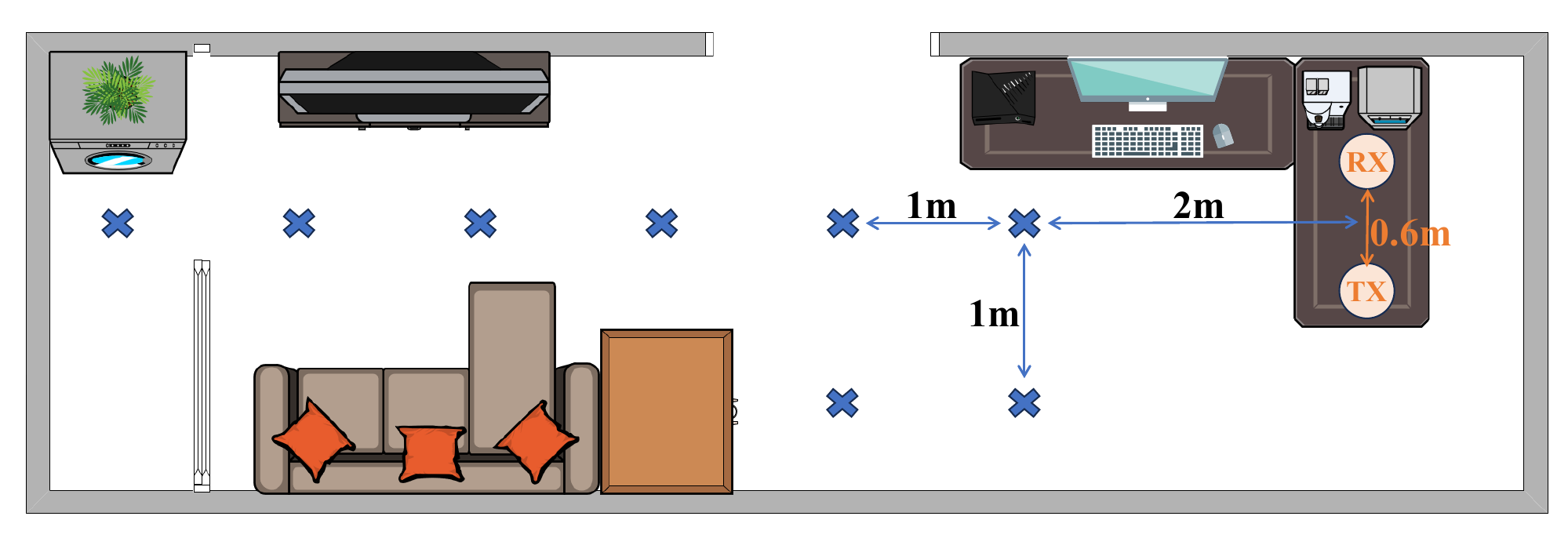}\label{fig:living}}
  \subfloat[Car park.]{\includegraphics[width=0.36\linewidth]{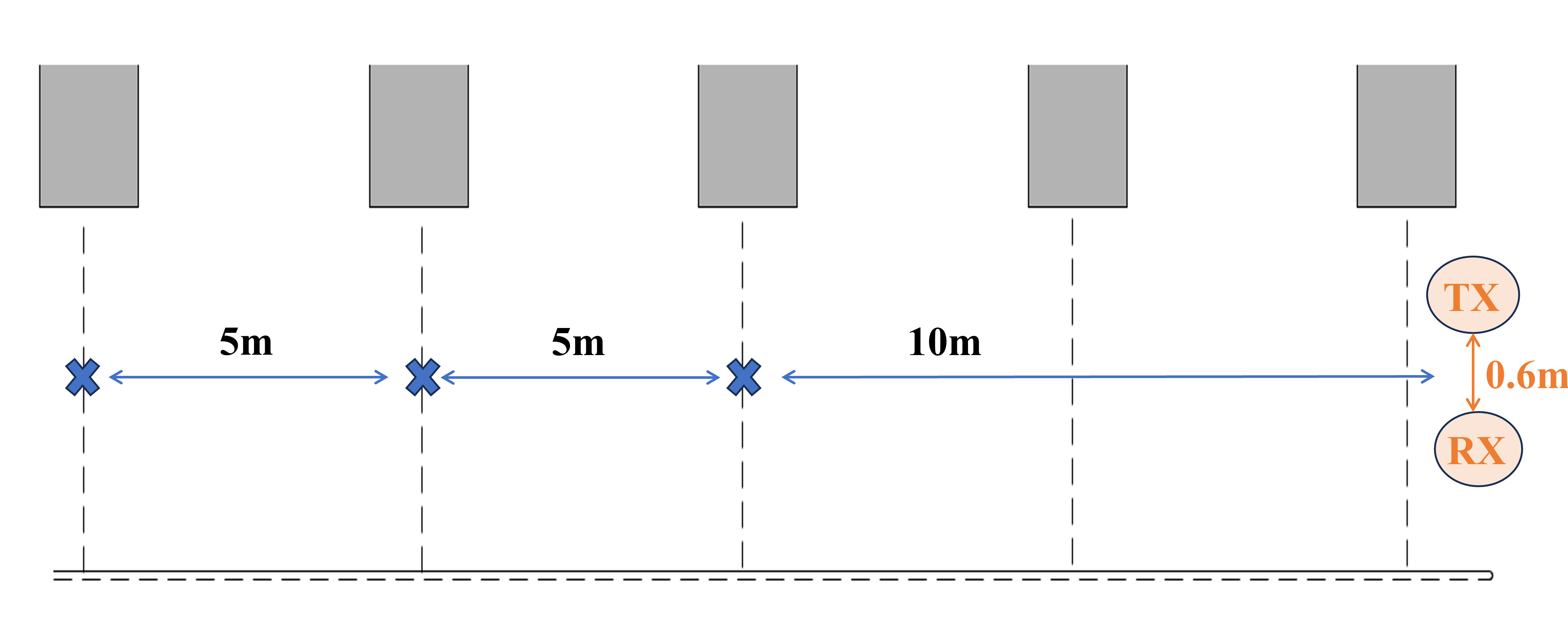}\label{fig:carpark}}
  \subfloat[Setup illustration.]{\includegraphics[width=0.15\linewidth]{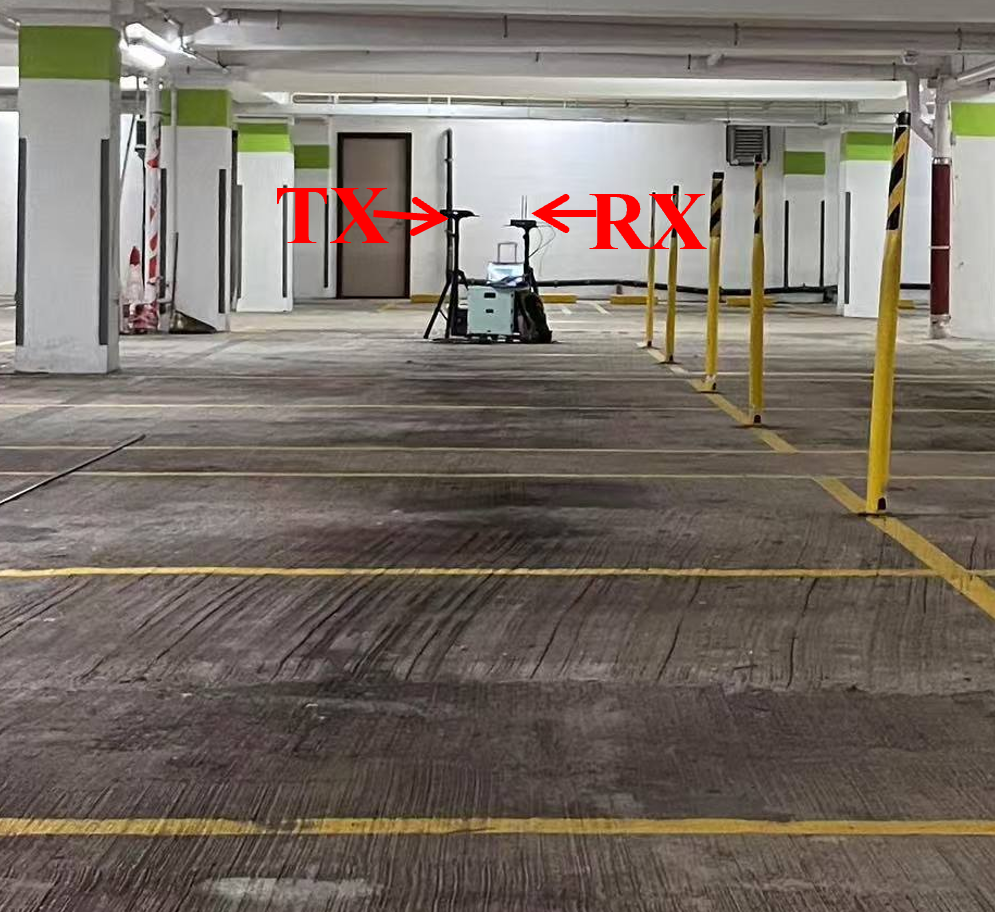}\label{fig:carpark_real}}
\caption{Experimental settings.}
   \end{figure*}
   
This experiment is designed to validate the efficiency of the proposed methods in maximizing the SSNR. Two distinct scenarios are designed to evaluate CIRSense: 
\begin{itemize}[leftmargin=*]
    \item A 2.7 m by 9 m real-life living room environment is used, as shown in Fig.~\ref{fig:living}. In this scenario, the transceivers are positioned 0.6 m apart, and the environment introduces complex multipath due to furniture, walls, and other obstacles. 
    \item The transceivers are placed 0.6 m apart in a car park, as illustrated in Fig.~\ref{fig:carpark} and~\ref{fig:carpark_real}. While car parks are not typical sensing environments, we intentionally selected this challenging scenario to rigorously evaluate far-range sensing capabilities, with target subjects positioned at considerable distances (10-20 m) away.
\end{itemize}

\begin{figure}
  \centering
  \subfloat[Distance estimation error]{\includegraphics[width=\linewidth]{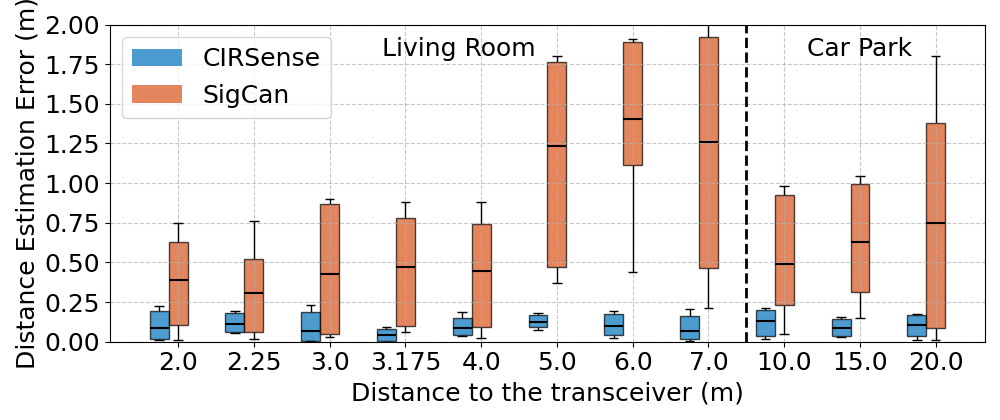}\label{tof_error}}
  
  \subfloat[Respiration rate error]{\includegraphics[width=\linewidth]{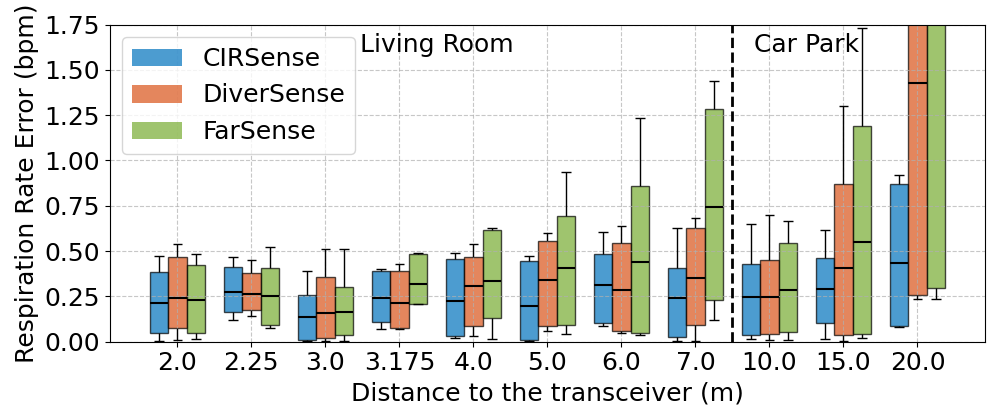}\label{bmp_error}}
\caption{Estimation errors under different distances using a single receiving antenna.}
\label{sensingrange}
 \vspace{-0.5cm}
\end{figure}

The performance comparison between CIRSense and baselines is conducted across various distances in both indoor and car park environments, as shown in Fig.~\ref{sensingrange}. To ensure rigorous statistical analysis, our results are presented using boxplots where the boundaries of the box body represent the 10th and 90th percentiles, capturing the core $80\%$ of the measurements. The whiskers extend to the minimum and maximum values, while the horizontal line within each box indicates the mean estimation error.

CIRSense demonstrates consistently superior performance across both living room (2-7 m range) and car park (10-20 m range) environments, significantly outperforming all baseline methods in both distance estimation and respiration monitoring applications. 
In the residential setting, CIRSense maintains exceptional distance estimation accuracy with maximal errors below 0.25 m and minimal variance, as shown in Fig~\ref{tof_error}. In comparison, Sigcan exhibits significantly higher mean errors and larger error variances within the same range. This performance gap becomes particularly pronounced at medium distances (5-7 m), where SigCan's estimation error increases dramatically while CIRSense maintains stable performance. This observation aligns with the findings reported in the original SigCan paper \cite{li2024sigcan}, where estimation results become unreliable beyond 5 m in indoor environments even using signals from three antennas. The inferior performance of SigCan at medium distances may be attributed to multipath effects. Similarly, for respiration monitoring, CIRSense maintains stable mean errors around 0.25 bpm with compact error distributions, while both FarSense and DiverSense demonstrate progressively degrading performance with increasing distance, with FarSense reaching mean errors near 0.75 bpm at 7 m distance.

The performance differential becomes even more pronounced in the challenging car park scenario. CIRSense maintains remarkable consistency in distance estimation with a mean error of just 0.11 m across the 10-20 m range, while SigCan exhibits significant performance degradation with errors exceeding 1 m at 15 m and 20 m distances. It should be noted that SigCan's performance in the car park environment, while still inferior to CIRSense, is better than in indoor scenarios despite the larger distances. This observation aligns with the original SigCan findings, where reduced multipath effects were shown to improve performance. However, the large error bars in SigCan's measurements indicate poor reliability and high measurement uncertainty in far-range sensing scenarios. The performance limitations of SigCan may stem from its reliance on phase linearity assumptions, which may not hold because of nonlinear phase distortion in practical RF chains \cite{liu2019real}. In contrast, CIRSense demonstrates superior robustness by operating without specific hardware assumptions, thereby achieving more reliable performance.

For respiration monitoring, CIRSense maintains errors below 0.5 bpm even at 20 m distance, while both baseline methods exhibit severe performance degradation, with DiverSense exceeding 1.3 bpm and FarSense reaching beyond 1.75 bpm at 20 m sensing distance. The superior performance of CIRSense over DiverSense and FarSense can be attributed to two key advantages. First, CIRSense demonstrates better signal quality in extracting respiration patterns, and second, it shows enhanced robustness against motion interference. By operating in the delay domain, CIRSense effectively separates target respiratory motions from disturbances through distinct delay tap concentrations, yielding clean sinusoidal waveforms that faithfully capture breathing cycles. This capability allows CIRSense to achieve reliable and accurate sensing across varied sensing distances and environments.

\subsection{Sensing in NLoS Settings}

\begin{figure}
    \centering    
    \subfloat[]{\includegraphics[width=0.5\linewidth]{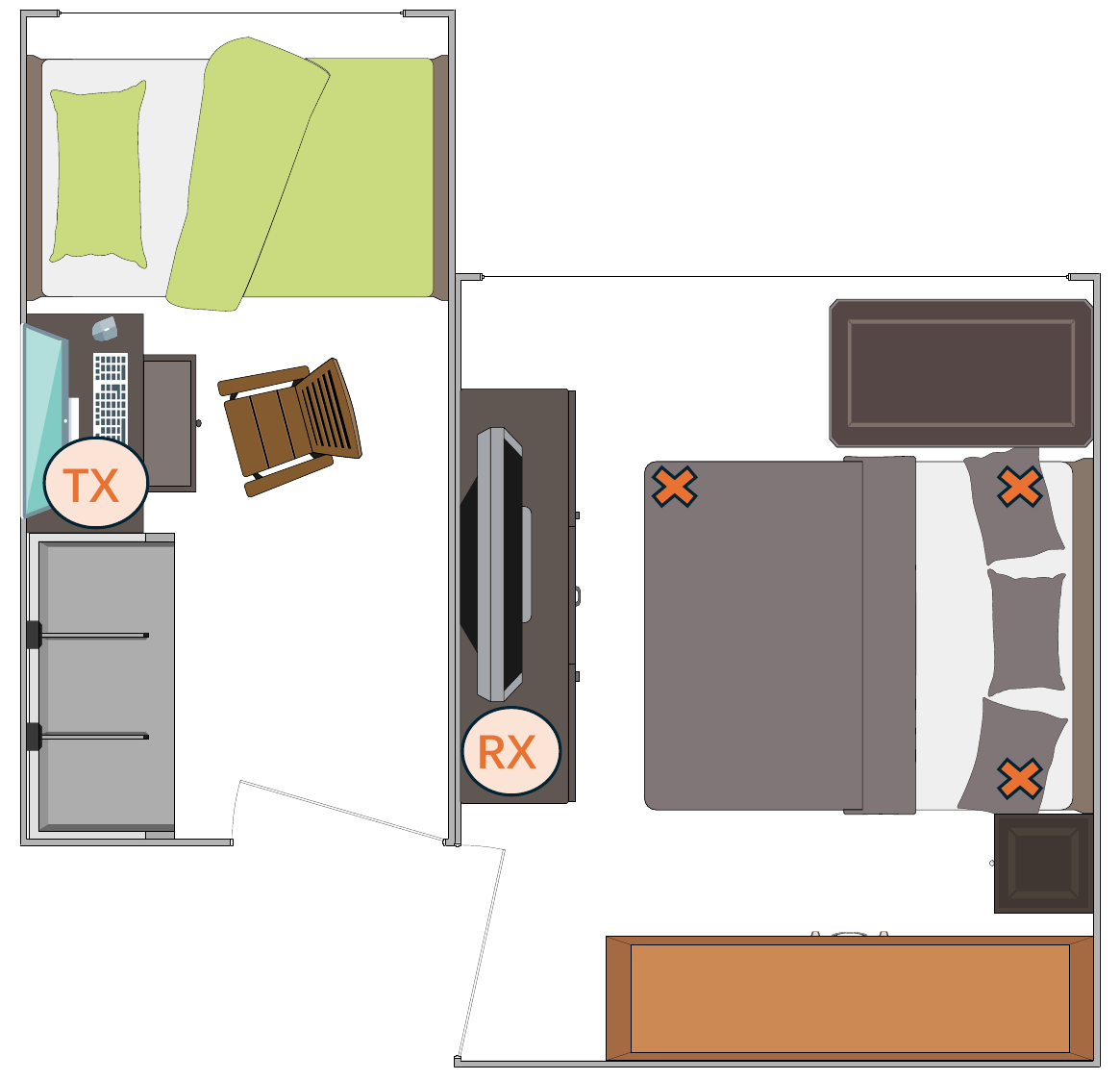}\label{fig:nlos_setup}}
  \subfloat[]{\includegraphics[width=0.5\linewidth]{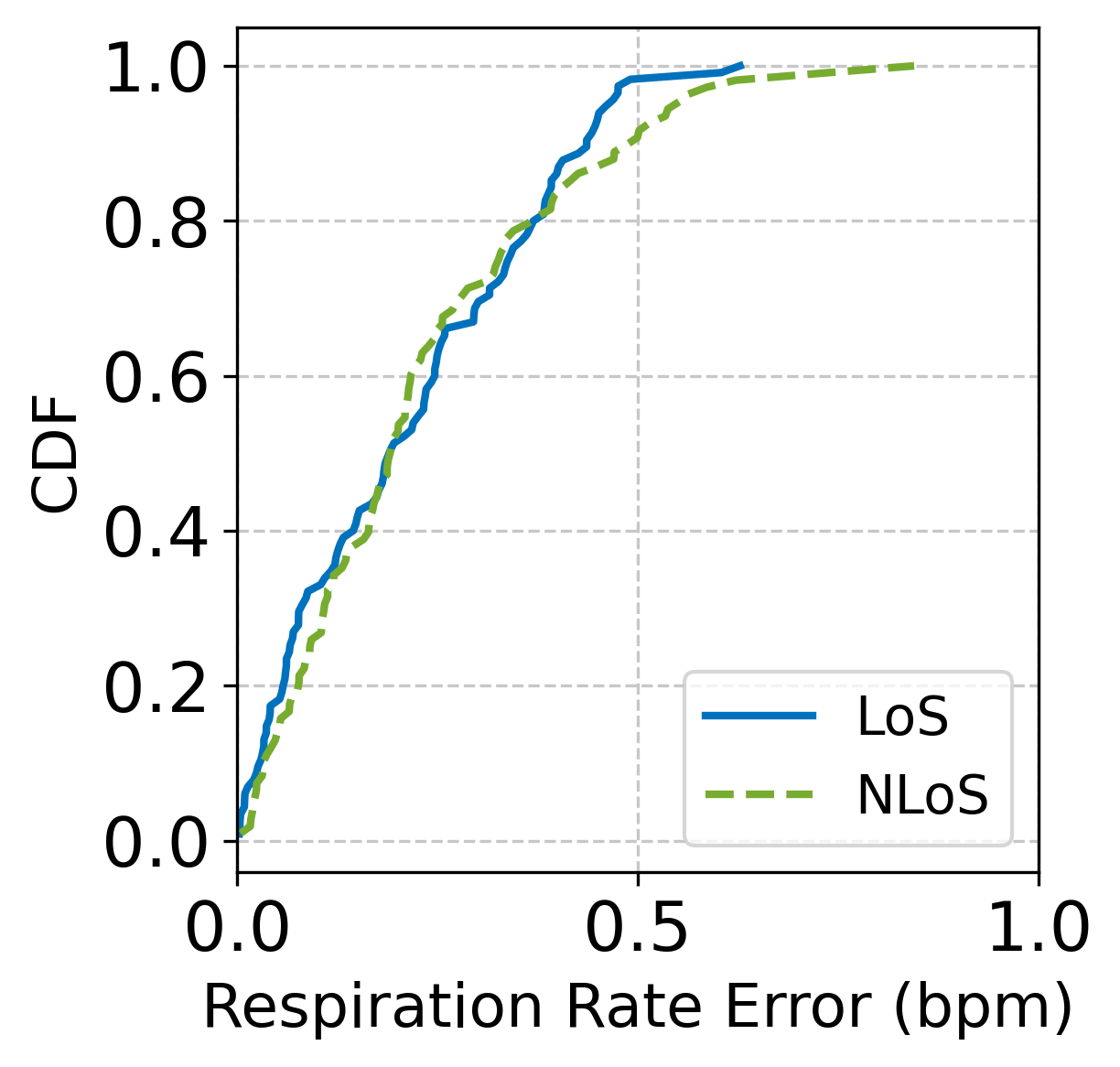}\label{fig:nlos_re}}
    \caption{Respiration sensing evaluation in NLoS settings. (a) Experimental setup in a residential apartment with two bedrooms. (b) CDF of respiration rate estimation errors.}
    \label{fig:nlos}
     \vspace{-0.5cm}
\end{figure}

This experiment is designed to validate the efficiency of our proposed method for RF distortion compensation in NLoS scenarios. As shown in Fig~\ref{fig:nlos_setup}, we conducted experiments in a typical residential apartment with two bedrooms. The transmitter (TX) was placed in one bedroom and the receiver (RX) located in the adjacent bedroom, creating a challenging NLoS environment. Red crosses in the figure represent the testing points where measurements were collected.

The cumulative distribution function (CDF) of respiration rate estimation errors is presented in Fig~\ref{fig:nlos_re}. 
The key finding from this experiment is that CIRSense maintain performance levels in NLoS conditions that are comparable to its performance in LoS scenarios in residential spaces. This indicates that the dominant static path remains a reliable reference for distortion compensation even when the direct path is obstructed by walls. Our CIRSense approach, which specifically leverages the dominant path identified through CIR analysis, demonstrates that effective distortion compensation can be achieved in NLoS environments without significant performance degradation. 
Notably, while CIRSense can estimate the total path length of the target-reflected path in NLoS settings, it remains challenging to retrieve the exact propagation path from transmitter to receiver. This is a common limitation shared by existing localization systems.

\subsection{Multi-Target Sensing}

\begin{figure}
  \centering
  \subfloat[Distance estimation error.]{\includegraphics[width=0.8\linewidth]{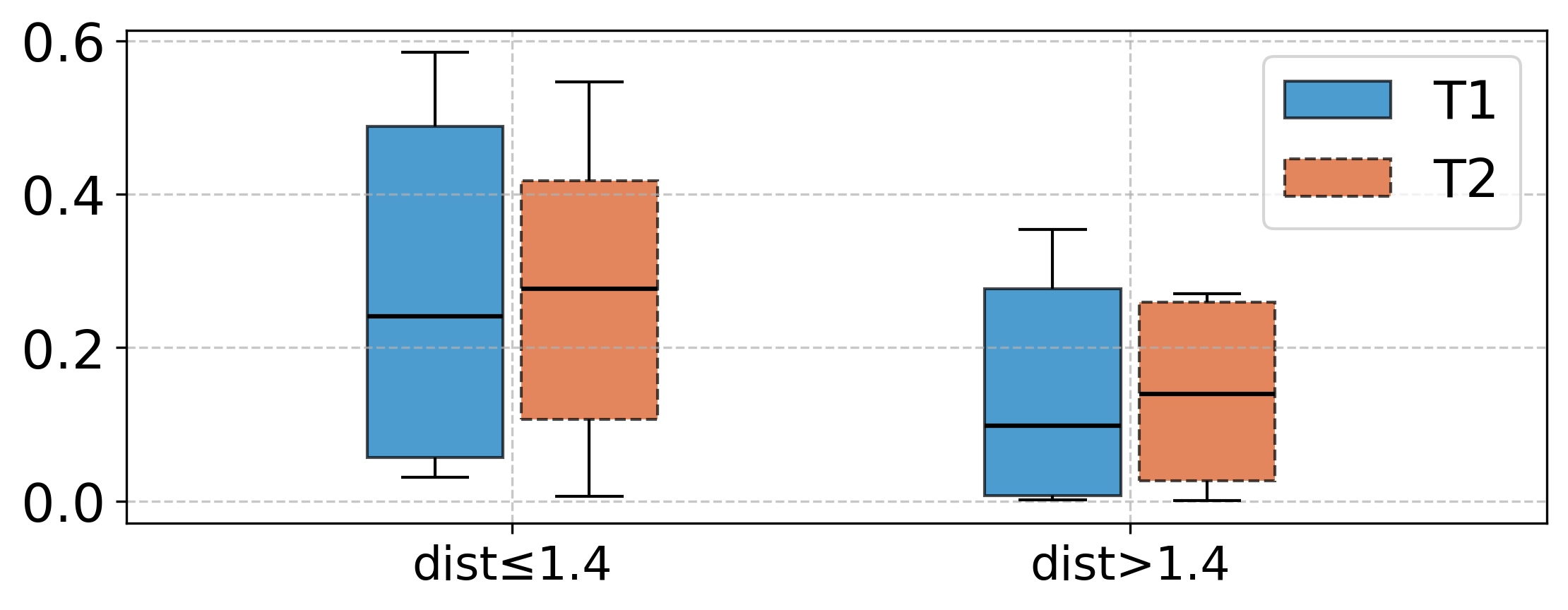}\label{fig:multi_d}}
  
  \subfloat[Respiration rate error.]{\includegraphics[width=0.8\linewidth]{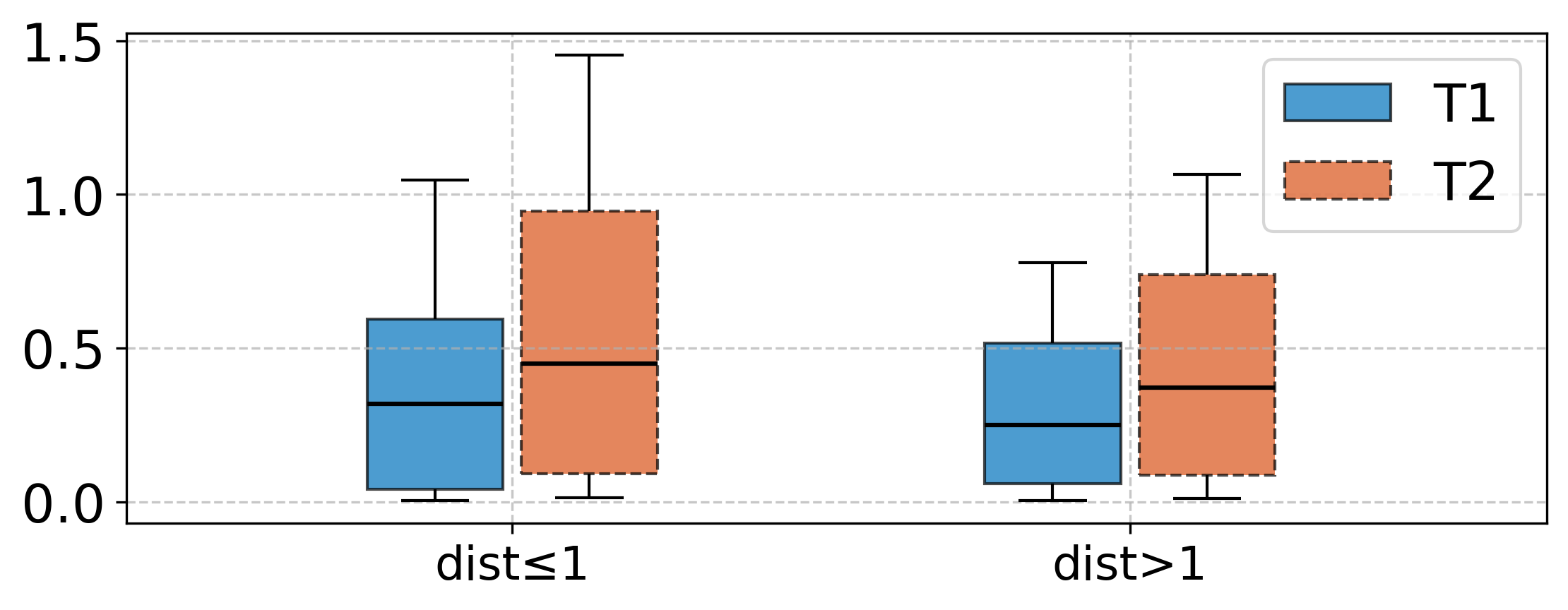}\label{fig:multi_b}}
\caption{Estimation error under multi-target scenarios, where dist represent inter-target distance.}
\label{fig:multi}
 \vspace{-0.5cm}
\end{figure}

In practical scenarios, multiple targets may coexist in the same environment, creating challenges. We evaluate the performance of our system in multi-target settings to assess its robustness in indoor scenarios. To our knowledge, our system is the first to demonstrate simultaneous multi-target respiration sensing and distance estimation using general commodity WiFi cards. Previous multi-target approaches have typically relied on specialized hardware \cite{karanam2019tracking, multiay23xiong}, very large bandwidth \cite{tan2019multitrack,multiay23xiong}, or multiple transceivers \cite{karanam2019tracking, hu2023muse}, limiting their practical deployment. We therefore report these results without baseline comparisons. Our multi-target sensing capability is achieved by identifying the first few taps with the highest variance as potential targets and sequentially applying our target path alignment process to each, prioritized by their variance magnitude.

Fig.~\ref{fig:multi} presents the estimation errors for both absolute target distance to the transceiver and respiration rate in environments with two targets (T1 representing the target closer to the transceiver than T2). We analyze performance under various target locations and varying levels of inter-target proximity. As shown in Fig.\ref{fig:multi_d} and Fig.\ref{fig:multi_b}, when targets maintain sufficient separation, our system achieves performance comparable to single-target scenarios for both tasks, with mean distance errors of approximately 0.1 meters for T1 and 0.14 meters for T2, and mean respiration rate errors of approximately 0.25 bpm for T1 and 0.37 bpm for T2. However, when targets are positioned in closer proximity, performance degrades for both metrics due to signal interference. This degradation occurs because the reflected signals from closely positioned targets create overlapping components in the CIR's delay taps, making it difficult for our system to differentiate between reflections from distinct targets. The movement from both subjects affect the changes in the same or adjacent taps, creating complex interference that complicates signal separation. Despite these challenges, the system maintains acceptable accuracy even in close-proximity scenarios, with mean errors increasing to about 0.24 meters and 0.28 meters for distance estimation (T1 and T2 respectively), and mean respiration rate errors remaining below 0.45 bpm for both targets.

\section{Discussions}

\subsection{Impact of Sampling Rate}

\begin{figure}
  \centering
  \subfloat[Distance estimation error.]{\includegraphics[width=0.9\linewidth]{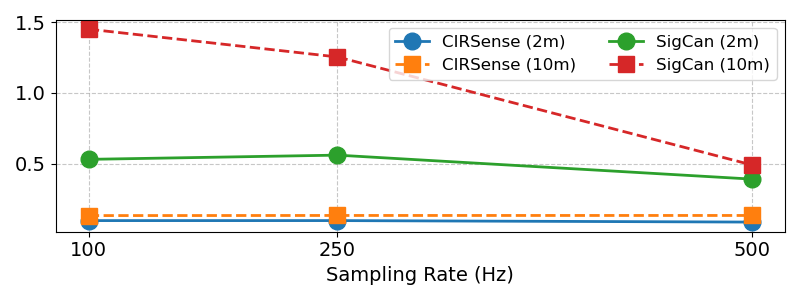}\label{fig:sr_d}}
  
  \subfloat[Respiration rate error.]{\includegraphics[width=0.9\linewidth]{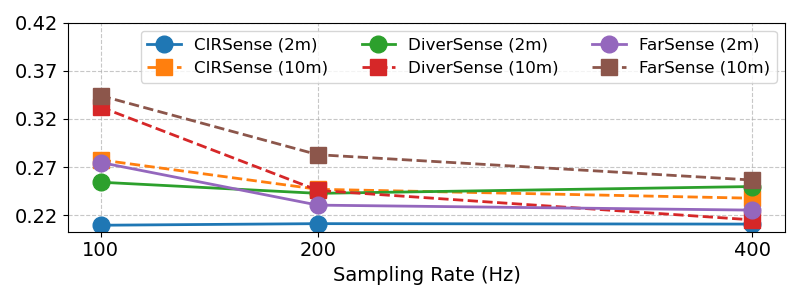}\label{fig:sr_b}}
\caption{Mean distance estimation error and mean respiration rate error under different sampling rates.}
\label{fig:sr}
 \vspace{-0.5cm}
\end{figure}

The influence of sampling rate on both distance estimation and respiration rate measurement accuracy is illustrated in Fig.~\ref{fig:sr}, comparing performance across different sensing methods and distances. Robust performance is maintained by CIRSense even at lower sampling rates, while higher dependencies on sampling frequency are observed in baselines.

As demonstrated in Fig.~\ref{fig:sr_d}, consistent distance estimation accuracy is achieved by CIRSense across all tested sampling rates (100-500 Hz), with mean errors maintained at approximately 0.09 m for both short (2 m) and long (10 m) ranges. In contrast, strong sampling rate dependence is exhibited by SigCan, particularly at extended distances. At 10 m, a large reduction in mean error is observed for SigCan, decreasing from 1.45 m at 100 Hz to 0.48 m at 500 Hz. This pronounced improvement suggests that reliable performance by SigCan requires a substantial number of samples. The superior low-rate performance of CIRSense may be attributed to its reliance on motion variance analysis, where sparse sampling proves sufficient for accurate characterization.

The effects of sampling rate on respiration rate estimation are presented in Fig.~\ref{fig:sr_b}. Similar stability across sampling rates is demonstrated by both CIRSense and DiverSense at 2 m distance, while a clear error reduction with increasing sampling rate is shown by FarSense. Notable improvements with higher sampling rates are observed for all methods at 10 m distance. For instance, at 10 m distance, DiverSense's error decreases from 0.34 bpm at 100 Hz to 0.21 bpm at 400 Hz. These improvements may result from both increased sample availability for noise reduction and enhanced temporal resolution of respiratory patterns. Despite these general trends, superior performance is maintained by CIRSense across all tested conditions, achieving the lowest mean estimation errors even at low sampling frequencies. Furthermore, minimal sampling rate dependence is exhibited by CIRSense, suggesting that effective denoising is accomplished with fewer samples through maximization of the SSNR at each sampling rate.

\subsection{Impact of RF Distortion Compensation}
\label{impact:dis_com}

\begin{figure}
  \centering
  \subfloat[Comparison combination.]{\includegraphics[width=0.7\linewidth]{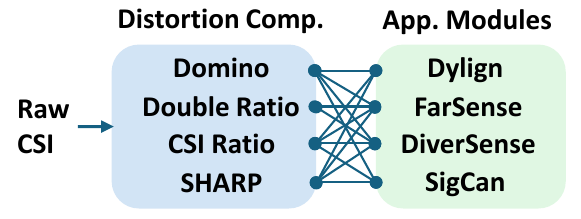}\label{fig:preprocess_c}}
  
  \subfloat[Distance estimation error.]{\includegraphics[width=0.9\linewidth]{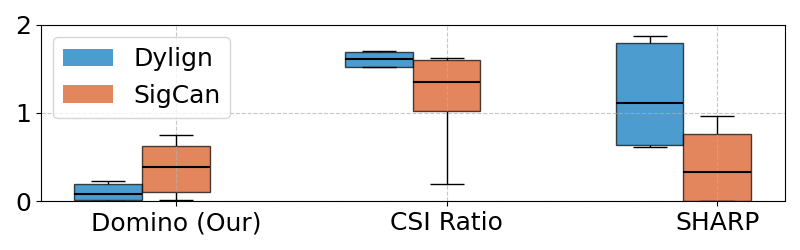}\label{fig:preprocess_d}}
  
  \subfloat[Respiration rate error]{\includegraphics[width=0.9\linewidth]{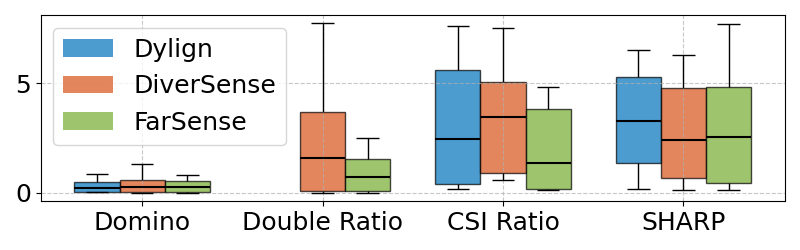}\label{fig:preprocess_b}}
\caption{Impact of RF distortion compensation methods.}
\label{fig:preprocess}
 \vspace{-0.5cm}
\end{figure}

Fig.~\ref{fig:preprocess} presents a comparative analysis of four distortion compensation approaches: our proposed Domino, Double Ratio \cite{enabling24yi}, CSI Ratio \cite{zeng2019farsense}, and SHARP \cite{meneghello2022sharp}. We compare sensing performance across different sensing application modules using data compensated by each compensation method, with the complete combination illustrated in Fig.\ref{fig:preprocess_c}. 
Note that the Double Ratio method destroys the physical structure of CSI with a reduced number of subcarriers during the compensation procedure, making it incompatible with our Dylign method and unsuitable for recovering target path parameters (e.g., delay). Due to this limitation, we only report its performance in respiration sensing with two baseline modules. 

As shown in Fig.~\ref{fig:preprocess_d}, we compare performance using data from 2 m scenarios. The proposed Domino achieves mean errors below 0.4 m with minimal variance for both Dylign and SigCan application modules, as evidenced by the compact boxplot distributions. In contrast, significantly larger errors and greater variability are observed for both CSI Ratio and SHARP methods. Specifically, mean errors exceeding 1.5 m are produced by CSI Ratio when applied to CIRSense, while substantial variance is introduced in SigCan method. SHARP's compensation method results in mean errors greater than 1.1 m for Dylign. Although slightly better performance is exhibited by SigCan with SHARP compensation compared to Dylign, considerable degradation relative to the proposed method is still observed. 
The inferior performance of CSI Ratio may be attributed to its fundamental dependence on the consistency of distortions between two receiving antennas. When this condition is not satisfied, significant errors are introduced. SHARP suffers from joint estimation errors in multipath components, leading to increased variability and limiting its application to large-scale human activity recognition \cite{meneghello2022sharp}. Furthermore, different fundamental dependencies are exhibited by SigCan and Dylign: SigCan relies on phase linearity of partial subcarriers in individual CSI measurements, while Dylign uses temporal variance analysis across all CIR measurements. Consequently, the algorithmic instability of SHARP is found to have a more pronounced impact on Dylign performance.

A more pronounced effect of distortion compensation is observed in respiration rate estimation using data collected under NLoS scenarios, as illustrated in Fig.~\ref{fig:preprocess_b}. Consistent performance is maintained by our Domino, with mean errors remaining below 0.281 bpm across all sensing approaches. In contrast, significantly degraded performance is exhibited by baselines, where error distributions are observed to extend up to 8 bpm. The large whiskers and box sizes in these cases indicate highly unstable measurements, limiting the practical applications of baselines. Even though Double Ratio achieves improved performance when paired with FarSense, our Dylign with Domino (CIRSense) still achieves 3× better mean accuracy.

\subsection{Processing Time Comparison}

\begin{figure}
  \centering
  \subfloat[Distortion compensation.]{\includegraphics[width=0.5\linewidth]{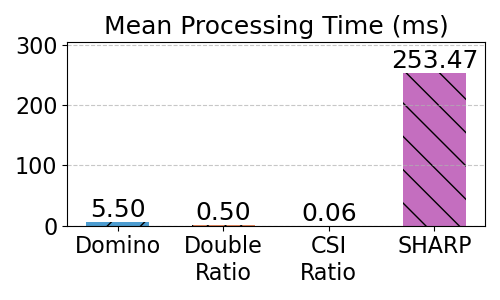}\label{fig:pt_p}}
  \subfloat[Distance estimation.]{\includegraphics[width=0.5\linewidth]{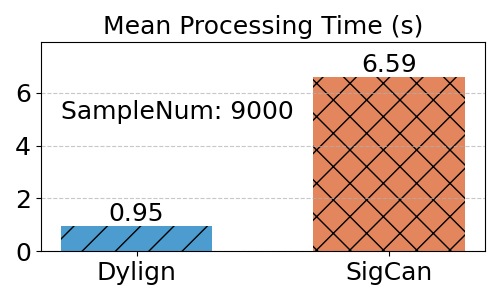}\label{fig:pt_d}}
  
  \subfloat[Respiration rate estimation]{\includegraphics[width=\linewidth]{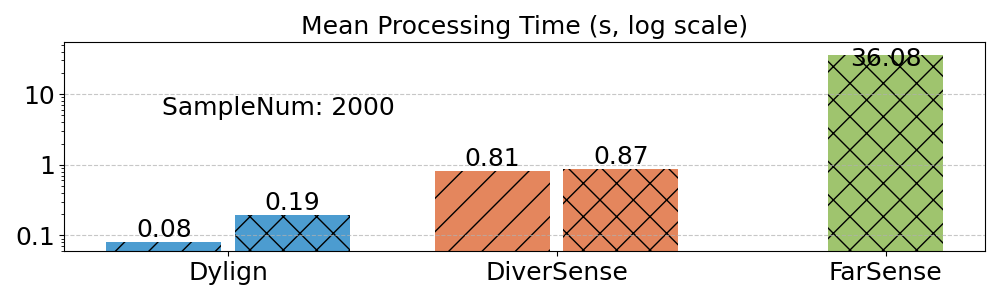}\label{fig:pt_b}}
\caption{Processing Time Comparison.}
\label{fig:pt}
 \vspace{-0.5cm}
\end{figure}

Fig.~\ref{fig:pt} presents a comprehensive evaluation of computational efficiency across different methods, examining processing times for three key aspects: RF distortion compensation, distance estimation, and respiration rate estimation. {All algorithms were implemented in MATLAB 2024b, with the exception of SHARP, which was developed in Python 3.12. To enhance computational efficiency, SHARP employs the OSQP solver \cite{osqp} to accelerate the solution of SHARP's core optimization problem. The experiments were conducted on a computer equipped with an Intel(R) Core(TM) Ultra 9 285H processor (2.90 GHz) and 32.0 GB of RAM.}

In RF distortion compensation (Fig.~\ref{fig:pt_p}), our proposed Domino requires an average of 5.50 ms to process a single CSI measurement, striking a balance between accuracy and efficiency. While Double Ratio and CSI Ratio show faster processing times at 0.5 and 0.06 ms respectively, their inferior compensation accuracy, as previously demonstrated in Fig.\ref{fig:preprocess}, makes this speed advantage less impactful. SHARP, despite its sophisticated approach, demands significantly more computational resources, taking 253.47 ms, nearly 46 times longer than that of Domino. This substantial processing overhead makes SHARP less suitable for practical applications.

For distance estimation (Fig.~\ref{fig:pt_d}), with a sample size of 9000 CSI measurements, Dylign demonstrates superior efficiency by completing calculations in 0.95 s. In contrast, SigCan requires 6.59 s for the same task, approximately 7 times longer than Dylign. This significant difference in processing time highlights Dylign's computational efficiency while maintaining higher accuracy.

The respiration rate estimation comparison (Fig.~\ref{fig:pt_b}) reveals interesting insights about processing efficiency with 2000 CSI measurements. {Dylign shows two critical processing time metrics: 0.08 s for dynamic path alignment and 0.19 s for total estimation time. Similarly, DiverSense requires 0.81 s for aligning sensing signals from all subcarriers and 0.87 s for total estimation time.} FarSense, operating with a different approach, requires 36.08 s for complete estimation processing. These results demonstrate that Dylign not only achieves faster overall processing but also maintains efficient SSNR maximization, requiring only about $10\%$ of the alignment time and about $22\%$ of total estimation time needed by DiverSense. The significant time difference between Dylign and FarSense (nearly 190 times faster) underscores Dylign's superior computational efficiency with better accuracy.

\section{Conclutions and Limitations}

The CIRSense framework presented in this work demonstrates advancements in WiFi sensing through principled exploitation of channel impulse response (CIR) properties. By shifting the sensing paradigm from conventional CSI-based approaches to delay-domain signal processing, we have successfully addressed three fundamental challenges in WiFi sensing: hardware-induced distortion, suboptimal subcarrier aggregation, and limited localization precision. The proposed CIR-based motion model provides a theoretical foundation for understanding target movement in the delay domain, while the practical implementation achieves robust performance through computationally efficient algorithms. Experimental results validate that CIRSense outperforms state-of-the-art methods, delivering around 0.25 bpm accuracy in respiration monitoring and around 0.09 m precision in distance estimation while maintaining at least $4.5\times$ higher computational efficiency, in diverse sensing scenarios. The demonstrated effectiveness of CIRSense in addressing core limitations of CSI-based approaches indicates that delay-domain processing represents a promising direction for future wireless sensing systems, particularly as WiFi technology evolves toward higher bandwidths and more advanced signal processing capabilities.

Several limitations warrant consideration for future research. First, while our framework supports multiple antennas, further exploration of advanced antenna diversity techniques could potentially yield additional performance improvements. Second, although our system demonstrates capability in multi-target sensing, performance degradation occurs when targets are positioned in close proximity, and challenges increase with more targets present, creating complex overlapping signal components. Developing more sophisticated algorithms could enhance multi-target resolution in these challenging scenarios. Third, our current approach shows promise in various indoor environments but lacks comprehensive NLoS localization capabilities that would require detailed room configuration information to accurately model signal propagation through obstacles. These limitations highlight promising directions for future work, including advanced spatial diversity techniques, enhanced algorithms for resolving closely positioned multiple targets, and context-aware NLoS localization methods that incorporate environmental mapping to account for complex indoor geometries.

 {\appendices
 	
 	\section{Preliminary Observations}
 	\label{pre_exp}

 	An experimental investigation was conducted in indoor scenarios involving a representative dynamic path of respiratory motion. The experimental setup consisted of transceivers spaced 60 cm apart, with a motion target positioned 5 m perpendicular to the LoS path (yielding a 9.4 m relative distance\footnote{Defined as the path length difference between dynamic and LoS paths}). CSI measurements were collected using the Picoscenes platform \cite{jiang2021eliminating} on a mini-PC with 160 MHz bandwidth. Additional implementation details are the same in Section~\ref{sec:exp_setting}. 
 	
 	Following effective RF distortion compensation via our reference path method, a direct comparison of SSNR is presented in Fig.~\ref{ssnr_com}, contrasting the highest-variance CSI subcarrier with the highest-variance CIR tap. The results clearly show that respiratory motion periodic patterns are more distinctly observable in CIR measurements, as CSI inherently distributes path power across multiple subcarriers. This finding underscores the power concentration capability of CIR.  
 	Furthermore, Fig.~\ref{dis_approx} illustrates the temporal variance distribution across CIR taps after distortion elimination. The results reveal that target motion predominantly influences a single CIR tap (tap 5 in this experimental configuration). By identifying this maximum-variance tap, the relative distance to the moving target can be estimated with high precision. The estimated relative distance can be calculated by $5\times T_s \times  c =9.375$m, showing only a $0.025$ estimation error. Here $c$ represents the speed of light. This observation further confirms the merits of our CIR-based sensing approach.
 	
 	\begin{figure}
 		\centering
 		
 		\subfloat[SSNR comparision]{\includegraphics[width=0.85\linewidth]{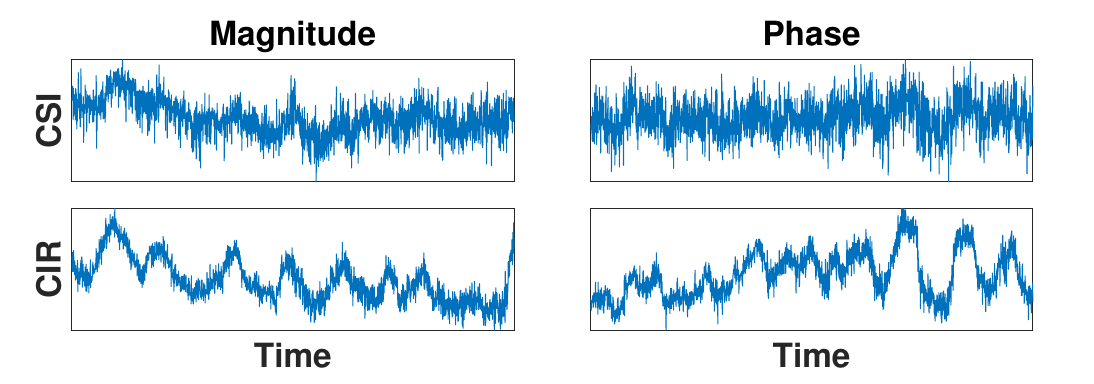}\label{ssnr_com}}
 		
 		\subfloat[Distance estimation]{\includegraphics[width=0.85\linewidth]{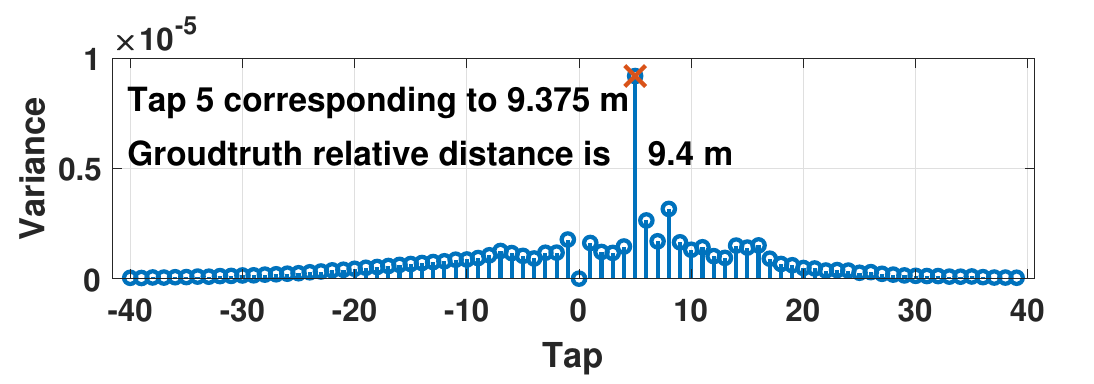}\label{dis_approx}}
 		
 		\caption{Preliminary observations.}\label{preliminary_result}
 		\vspace{-0.5cm}
 		\label{benefits}
 	\end{figure}
 	
	\section{CIR-based Sensing Model Validation}
	\label{modelvalidation}
	
	In this appendix, we intend to validate the proposed motion model for CIR-based WiFi sensing through controlled experiments. 
	In these experiments, the Picoscenes platform \cite{jiang2021eliminating} is employed for CSI measurements collection, with the WiFi transceiver configured to operate at 5.25 GHz and 160 MHz bandwidth, see more details in Section~\ref{sec:exp_setting}. Time-varying RF distortions are first mitigated using the methods outlined in Section~\ref{rfdistortion_cirsense}, enabling accurate CIR acquisition from CSI measurements. The transceivers are positioned 30 cm apart, and a programmable linear motion slider with a mounted metal plate, as shown in Fig.~\ref{mv_setup}, is used to emulate precise distance displacement perpendicular to the LoS path. The metal plate is programmed to move linearly from around 1.09 m to 1.33 m over a duration of 3.2 seconds, corresponding to an initial position of $T_s\times c$ relative to the LoS path and a total path length displacement of 8.5 wavelengths $\lambda$ (equivalent to 0.486 m in this configuration).  
	
	As part of the data pre-processing, the estimated static component, characterized by the mean value of each tap, is subtracted from the CIR measurements. A moving average mechanism is then applied to smooth the complex-valued CIR, enhancing the visualization of trends. Fig.~\ref{mv} illustrates the trajectory of the value of the second tap $h[1]$, which is the tap with the maximal variance and corresponding to a delay of $T_s$. {Specifically, the blue and red points mark the initial and final tap values, respectively. In this experiment, the initial relative path length of the target is exactly $T_s$, resulting in ideal sampling at tap $h[1]$. As the metal plate moves away from the transceiver, the increasing relative path length exceeds $T_s$, introducing fractional delays due to non-integer sampling at tap $h[1]$.} 
	We observe that the second tap reveals distinctive circular patterns in the complex plane during plate motion. Specifically, the trajectory completes 8.5 rotations, with the rotation radius decreasing due to the changing values of  $p[n,\tau[t]]$.\footnote{The decreasing rate is determined by the characteristics of the pulse shaping filter used in the system, the environmental attenuation factor, and reflection attenuation.}  
	This observed behavior coincides with the theoretical analysis for distance displacements, as illustrated in Fig.~\ref{fig:motion_model}. In respiratory monitoring applications, sub-wavelength chest motion manifests in the CIR through characteristic circular trajectories in the complex plane, a phenomenon primarily attributable to phase variations induced by small-scale movements. Additionally, it is observed that the target’s spatial position influences the power of the target-reflected signals, which further impacts the accuracy of respiration sensing.
	
	\begin{figure}
		\centering
		\subfloat[]{\includegraphics[width=0.8\linewidth]{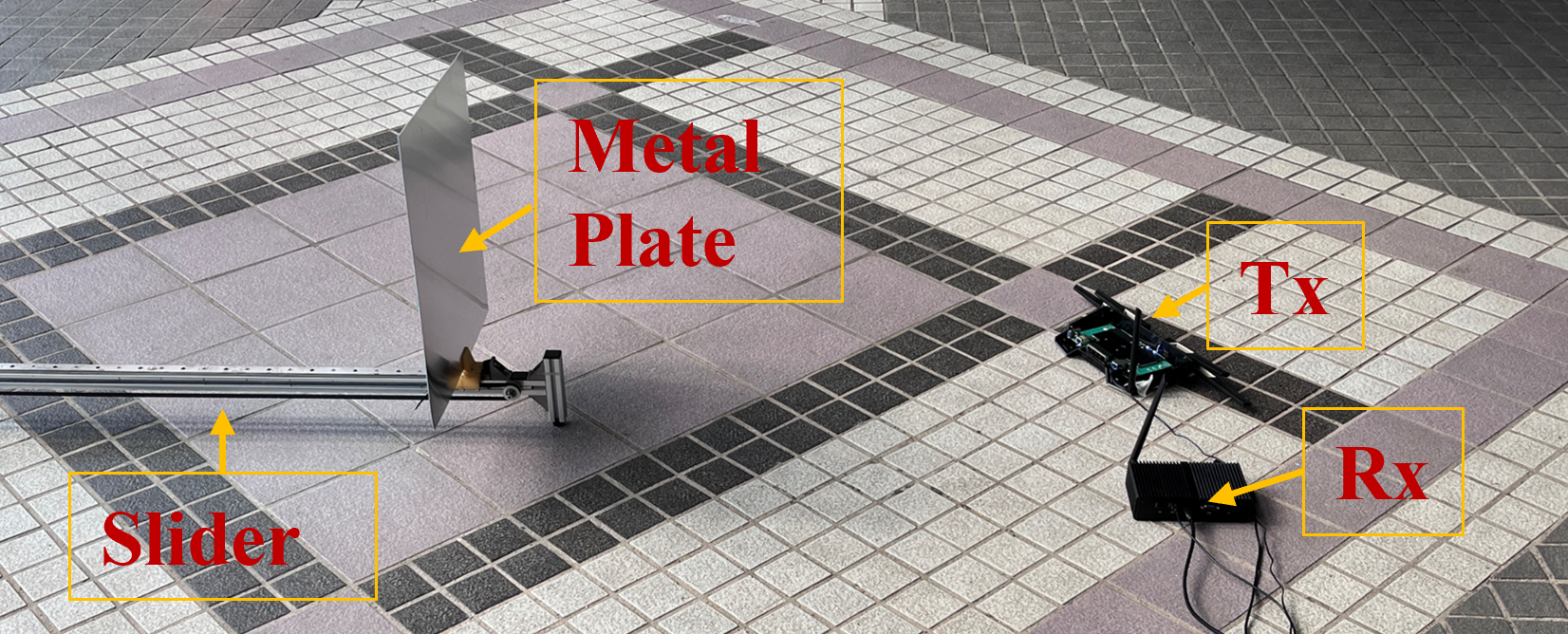}\label{mv_setup}}
		
		\subfloat[]{\includegraphics[width=0.7\linewidth]{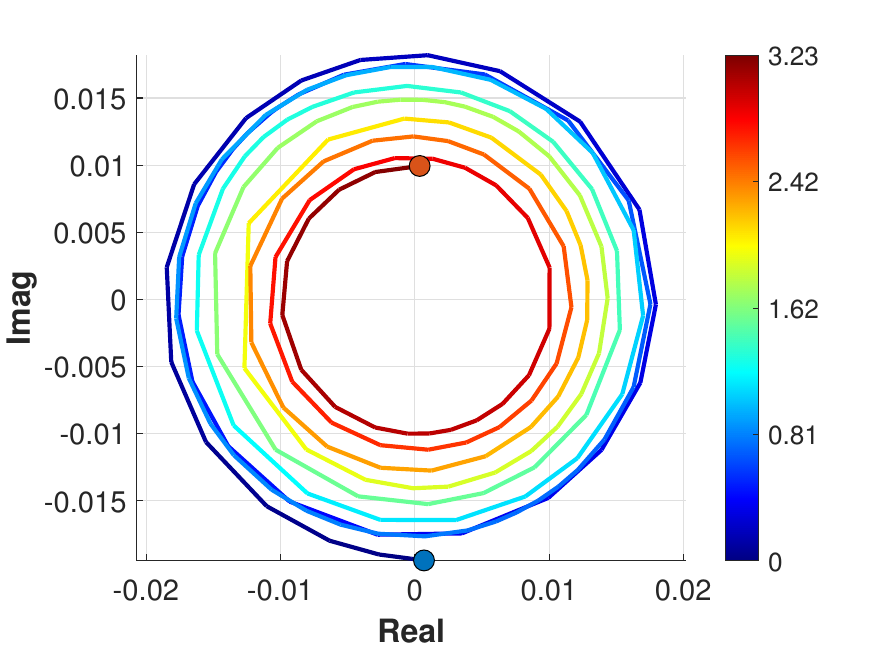}\label{mv}}
		\caption{Experimental verification for CIR-based motion model. (a) Experiment setup. (b) Complex plane trajectory of the target tap in the CIR.}
		\vspace{-0.5cm}
	\end{figure}
	
	\section{CIR Acquisition}
	\label{cir_acq_val}
	
	\begin{figure}
		\centering
		\includegraphics[width=0.8\linewidth]{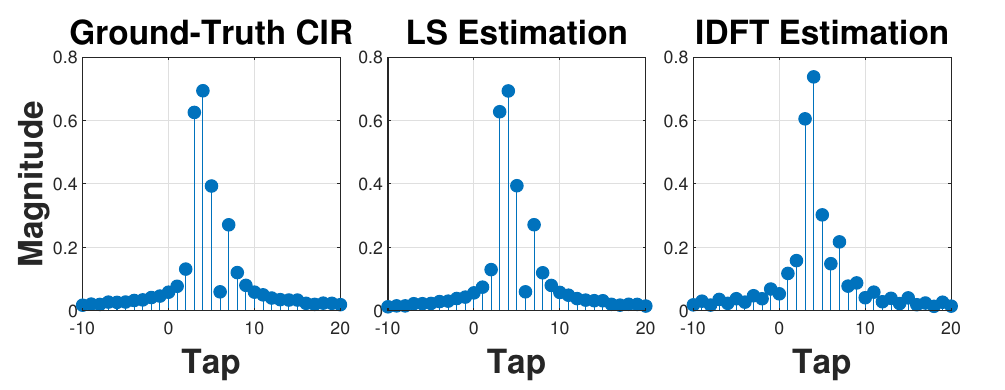}
		\caption{CIR acquisition based on LS and IDFT estimation.}
		\label{fig:cir_est}
		\vspace{-0.5cm}
	\end{figure}
	
	A simulation-based approach is adopted to validate the LS method proposed for CIR acquisition, given the inherent difficulties in obtaining ground-truth channel information in practical scenarios. The comparative results presented in Fig.~\ref{fig:cir_est} reveals the effectiveness of the LS estimation method for CIR acquisition from CSI. As illustrated in the middle subfigure, the LS approach generates tap estimates that closely approximate the ground-truth CIR shown in the left subfigure, particularly in maintaining accurate tap magnitude characteristics. In contrast, the direct IDFT-based estimation results, shown in the right subfigure, demonstrate noticeable deviations from the ground truth. 
	
	\section{Implementation of Fractional Delay Compensation} 
	\label{principle_fractionaldelaycom}

	{The delay-shifting operation can be implemented efficiently in the frequency domain by applying phase shifts to the CSI measurements. This approach assumes that the delay shift within the pulse function $p(\cdot)$ primarily introduces phase shifts while preserving the magnitude of its frequency response. This assumption is valid due to two reasons. First, the combined effect of maximal path delay and RF distortion-induced delay offset must remain within the cyclic prefix duration, which is shorter than the OFDM symbol length \cite{80211ac}. This constraint ensures the channel impulse response remains confined to the central taps, with other tap coefficients vanishing except for noise components. Under these circumstances, the applied delay shift effectively becomes a cyclic shift operation. 
		Second, fundamental DFT properties dictate that such cyclic shifts in the delay domain induce linear phase shifts across subcarriers in the frequency domain. Therefore, a delay shift $ \Delta' $ in the CIR is equivalent to multiplying the CSI by a complex exponential $ e^{j 2\pi k \Delta'/N} $, where $ k $ is the subcarrier index and $ N $ is the total number of subcarriers.} This approach allows the fractional delay to be corrected without requiring additional precise pulse-shaping function values for interpolation.

	 }

\bibliographystyle{IEEEtran}
\bibliography{ref}

\end{document}